# Likelihood Ratio as Weight of Forensic Evidence:
## A Metrological Perspective

Steven P. Lund  
steven.lund@nist.gov

Hari Iyer  
hari@nist.gov

Statistical Engineering Division, Information Technology Laboratory,  
National Institute of Standards and Technology, Gaithersburg, MD 20899, USA

## Abstract

Lindley (1977) laid the groundwork for a statistical treatment of evaluating the weight of evidence in forensic science based on a subjective Bayesian formulation of the problem. He noted that the decision maker (DM) need only multiply her likelihood ratio ($LR$) for the evidence (Lindley simply called it *the factor multiplying odds*) by her prior odds for guilt/innocence of the defendant to obtain her posterior odds for the same. The current thinking among many leading forensic scientists and statisticians engaged in forensic evidence interpretation is that a forensic expert should compute a $LR$ as a summary of their analysis and present its value as the weight of evidence to interested parties in a written report or to the triers-of-fact during testimony (see ENFSI (2015)). The DMs could then use the $LR$ value provided by the forensic expert to update their own respective prior odds of guilt to arrive at their posterior odds. Presenting a $LR$ value as the weight of evidence is seen as an application of Bayesian decision theory, and, by mathematical arguments of coherence and rationality, is therefore normative (*i.e.*, forensic experts *should* communicate in this manner).

In general, forensic experts are called upon not to reach actionable decisions, but to help inform others who are charged with making a decision. How to decide something for yourself and how to relay information to others are two starkly different questions studied under separate theoretical disciplines. Transferring information is not within the purview of decision theory which considers only the decision making entity. Facilitating a meaningful exchange of scientific observations among interested parties is a fundamental purpose of measurement science and the suitability of reporting a $LR$ must therefore be assessed in a very different manner.

From the perspective of metrology, the proposed $LR$ framework asks the forensic expert to measure the weight of evidence on behalf of the DM and report its value for subsequent use in Bayes formula. In this context, an appropriate uncertainty statement would assess the extent to which the expert's $LR$ value may differ from the $LR$ perceived by any given DM following careful review of the complete body of evidence considered by the expert. While difficult, or maybe even impractical, this assessment is nonetheless critical to establishing the suitability of transferring a $LR$ value as the weight of forensic evidence. Overlooking or dismissing the relevant uncertainty treats the value obtained by an expert as though it is a universal and exactly accurate characterization of weight of evidence - a perfect measurement. Given the degree to which any $LR$ depends upon subjective beliefs and choices of its evaluator, we are concerned that assessments of repeatability, reproducibility, and traceability, properties that give a quantity the status of a measurement, would find that $LR$ values are not sufficiently transferable for their intended use.

In this article we provide a rebuttal against the possible perception that a single number, such as the $LR$, can provide an objective or definitive weight of evidence. We argue that presenting a probabilistic interpretation of evidence in the form of a $LR$ would require an extensive uncertainty characterization. We illustrate the concepts of a *Lattice of Assumptions* and an *Uncertainty Pyramid* as tools for evaluating the uncertainty attributable to the choice of assumptions used during analysis under the constraint that any assumptions are judged to be consistent with available empirical information. When such analyses are considered untenable, we argue the $LR$ simply should not be considered amenable to measurement by one person on behalf of another and alternative presentations of evidence should be pursued. In our view, forensic experts should not feel it is their duty to provide a steadfast interpretation. Rather, a forensic expert should assist DMs form their own interpretations from a clear understanding of the relevant, objective, and demonstrable information. We hope this article will help clarify distinctions between personal interpretations and transferable information and inspire those in the forensic science community to carefully consider principles of measurement science when establishing their professional communication practices.

# Introduction

In criminal and civil cases alike, the judicial system involves many individuals making decisions following exposure to some form of evidence (*e.g.*, district attorneys deciding whether or not to file criminal charges, prosecution or defense attorneys deciding or advising their clients whether to accept a plea agreement or proceed to trial, jurors voting guilty or not guilty). These decision makers (DMs) often rely on the findings of forensic experts (and, in the case of attorneys, how those findings will be perceived by others), whether expressed as a written report or through testimony at a trial, to help inform their decision. How experts express their findings and how DMs factor that information into their ultimate decisions remain areas of great public importance and current research; see, for example, Thompson and Newman (2015) and Thompson, *et al.*, (2013).

Lindley (1977) laid the groundwork for a statistical treatment of evaluating the weight of evidence[1] in forensic science.[2] His framework is based on a subjective Bayesian modeling of the problem which leads to the equation

$$\text{Posterior Odds}_{\text{DM}} = \text{Prior Odds}_{\text{DM}} \times LR_{\text{DM}}.$$

This odds form of the Bayes rule appears to separate the ultimate degree of doubt a DM feels regarding the guilt of a defendant, as expressed via posterior odds (*i.e.*, probability of guilt after considering the evidence divided by one minus this probability), into probability of guilt before consideration of the evidence at hand (prior odds) and the influence or weight of the evidence (expressed as a likelihood ratio).

Though Lindley (2014) makes it clear that the $LR$ in Bayes formula is the personal $LR$ of the DM, some scholars have suggested treating $LR$ as a quantity that can be (and should be) calculated by a forensic expert; see, for instance, (Aitken and Taroni (2004), chapter 3; ENFSI, 2015). The forensic expert can then communicate the result to the DMs[3], who could then apply (or envision others applying) Bayes rule to modify their respective prior odds by the reported $LR$ and arrive at their posterior odds as to the guilt/innocence of the defendant. This proposed adaptation can be expressed by the equation

$$\text{Posterior Odds}_{\text{DM}} = \text{Prior Odds}_{\text{DM}} \times LR_{\text{Expert}}.$$

DMs then choose their actions (*e.g.*, a district attorney decides to file criminal charges, a juror decides to vote not guilty, etc.) according to their perceived consequences of the actions, given the resulting posterior probabilities.

The appeal of this adaptation of Lindley's approach is that an impartial expert examiner could determine the meaning of the evidence by computing a likelihood ratio ($LR$), while avoiding strictly subjective initial perspectives regarding the guilt or innocence of the defendant. This hybrid approach is embraced by many forensic scientists in several European countries and is currently being evaluated as a candidate framework for adoption in the United States.

---

[1] The term *weight of evidence* appears in the book *Probability and Weighing of Evidence* by I. J. Good (1950), much earlier than Lindley's *Biometrika* (1977) paper.

[2] For a general exposure to the potential role of probability and statistics in the law the reader may consult Fienberg (1989), Dawid (2002), and Kaye and Freedman (2011).

[3] ENFSI (2015) guidance document provides several examples illustrating how forensic examiners may use subjective probabilities to arrive at a $LR$ value to convey to the DMs the strength of evidence they examined. Furthermore, this guidance document also indicates that forensic examiners may convert the numerical $LR$ value into a verbal equivalent following some scale of conclusions. Verbal expressions, however, cannot be multiplied by prior odds to obtain posterior odds.

Lindley's original framework provides a rational and coherent[4] approach for an individual to quantify her personal weight of forensic evidence. One of its greatest contributions is the clear articulation that the meaning of the evidence cannot be established solely through the comparison between the crime scene sample and a test sample. Lindley's $LR$ framework highlights the importance of the role of alternative explanations for the collection of observed evidence (see Lindley (2014), pages 117, 128), which, due to the sheer multitude of possibilities, will likely require greater effort to investigate than will the prosecution claim alone. Additionally, when the required personal components for computation are known for a given individual, Lindley's $LR$ provides a straightforward path for quantifying that individual's precise weight of evidence on a meaningful and standardized continuum.

The general steps required for such an evaluation are as follows.

- The DM constructs a collection of all explanations to be considered for how the observed evidence could have originated. Constructing a $LR$ requires partitioning this collection of considered explanations into two sets.[5] Suppose the DM is a juror who will cast a vote of either 'guilty' or 'not guilty' at the conclusion of a trial. The DM may assign any considered explanation for the observation of the evidence to one of two categories, *guilty* and *not guilty*, according to whether he/she would declare the defendant to be *guilty* or *not guilty* if the given explanation were known to be exactly true.

  Let $H_{p1}, H_{p2}, \ldots, H_{pa}$ denote the mutually exclusive and exhaustive collection of explanations under each of which the DM would declare the defendant to be guilty. We have

  $$H_p = H_{p1} \cup H_{p2} \cup \ldots \cup H_{pa}.$$

  Likewise, we write $H_{d1}, H_{d2}, \ldots, H_{db}$ for the mutually exclusive and exhaustive collection of explanations under each of which the DM would declare the defendant to be not guilty. We have

  $$H_d = H_{d1} \cup H_{d2} \cup \ldots \cup H_{db}.$$

- After exhaustively sorting the set of considered explanations, the DM assigns his/her (prior) degree of belief in each explanation before considering the evidence itself. This is done by assigning a non-negative (subjective) probability to each explanation such that the sum of all the probabilities is one. Let $Pr[H_{pi}]$ be denoted by $\pi_{pi}$ and $Pr[H_{dj}]$ be denoted by $\pi_{dj}$. Denote the sum $\sum_{i=1}^{a} \pi_{pi}$ by $\pi_0$. Then the sum $\sum_{j=1}^{b} \pi_{dj}$ equals $1 - \pi_0$. Here $\pi_0$ is the prior probability from the perspective of the DM that the defendant is *guilty* and $1 - \pi_0$ is the corresponding prior probability that the defendant is *not guilty*. We write $w_{pi} = \dfrac{\pi_{pi}}{\pi_0}$ and $w_{dj} = \dfrac{\pi_{dj}}{1 - \pi_0}$. Thus $w_{pi} = Pr[H_{pi}|H_p]$ and $w_{dj} = Pr[H_{dj}|H_d]$. Note that any explanation not explicitly given a positive prior weight, including those that the DM has accidentally overlooked[6], is given a prior weight of zero.[7]

---

[4] For systematic introduction to the statistical meanings of "rational" and "coherent," see Lindley (2004).

[5] The authors note that this constraint may prevent the $DM$ from maximizing expected utility, as Bayesian decision theory advises, when the utility assigned to a given action varies across explanations within one or both sets.

[6] If at any point the DM becomes aware of a previously overlooked explanation that will be given a prior weight greater than zero, the DM would assign the new explanation to its appropriate group, renormalize the weights within that group so that they sum to one, and continue the likelihood ratio calculation.

[7] A prior weight of zero indicates that the DM would never consider the explanation as plausible regardless of what data were presented. This hardline stance would seem more likely to be taken unintentionally or as a matter of convenience rather than conviction. That is, the entire collection of explanations with an assigned prior weight of zero can be removed from further consideration, greatly reducing downstream workload. Additionally, even the most outlandish explanations could become seemingly irrefutable, provided sufficient data. By this notion, it seems unlikely that any prior probability is rigid and exactly zero.

- For each explanation with prior weight greater than zero, the DM is to assess the probability of the observed evidence $E$ occurring among all outcomes that could result from the described scenario. Let $L_{pi}$ denote the likelihood of observing the evidence under explanation $H_{pi}$. (Some may find it more natural to denote this quantity as $Pr[E|H_{pi}]$, but we will use $L_{pi}$ for succinctness.) Similarly, let $L_{dj}$ denote the likelihood of observing the evidence under explanation $H_{dj}$.
- Once a weight and a likelihood have been determined for each explanation of the observed evidence, the likelihood ratio is given as the sum of the products of the likelihood and the corresponding prior weight[8] for each explanation in the *guilty set* divided by the sum of the products of the likelihood and the corresponding prior weight for each explanation in the *not guilty set*. This may be expressed algebraically as follows:

$$LR = \frac{\sum_{i=1}^{a} L_{pi} w_{pi}}{\sum_{j=1}^{b} L_{dj} w_{dj}}. \qquad (1)$$

The $LR$ is insensitive to the redistribution of prior weights among explanations that share a common likelihood within the *guilty set* (or within the *not guilty set*). In the context of source attribution, for instance, the DM may believe the alternative sources are a random sample from a particular population and not have any additional information that would lead to assigning different likelihoods among the alternative sources. In this instance the DM might assign each alternative source a likelihood representing the probability of observing the evidence by random selection from that population, and the denominator becomes that same probability, regardless of what weights $w_{dj}$ would be chosen.

**Practical Considerations**

Computing a $LR$ for anything but the simplest of problems involves approximations. Rather than assign prior weights that exactly and genuinely reflect one's personal belief, tractable and familiar substitutions are made. One hopes that the resulting value is not overly sensitive to such choices, but a rigorous uncertainty analysis is often difficult and not pursued. In such instances the computed value can be viewed either as an approximation of unknown accuracy for the ratio between posterior and prior odds; or as a *score* for which the appropriate interpretation becomes unclear under Bayesian decision theory, even for the person who has computed it. Of course, any individual need only satisfy herself regarding the perceived suitability of the computed $LR$ for personal use in Bayes formula. Attempting to compute a probabilistic interpretation on behalf of another involves greater responsibility.

The $LR$ in Lindley's framework is, by definition, personal and subjective, which does not prompt an expectation of transferability from one person to another. Some analysts may refer to a collection of demonstrable data and/or theoretical arguments to suggest that a particular probabilistic interpretation is normative rather than subjective. However, personal choice often pervades these arguments as well: presuming representativeness of data not sampled according to an established statistical method; defining the plausibility of candidate models in light of available data (*e.g.*, binary inclusion/exclusion assignments based on coverage intervals under frequentist inference, posterior distributions under Bayesian inference); supposing a finite sample behaves in accordance with assumptions and asymptotic theory. In lieu of strict and demonstrably necessary constraints leading all parties to a particular analysis result, personal choices are made.

---

[8]These weights assigned within each category illustrate that even computing a $LR$ is not free from prior probability assignment at the level of specific explanations.

Perhaps in recognition of the subjectivity of any given interpretation, stating one's assumptions is a core component of a statistical report. This is necessary, but not sufficient. Stating assumptions helps facilitate transparency, but not transferability. Transparency enables a trained audience to label a presented analysis as reasonable or unreasonable - much like a statistical hypothesis test. Of greater relevence for transferability is the set of results corresponding to all analyses that are seemingly plausible (the analog of a statistical confidence interval). Completely dismissing the importance of uncertainty in this context is equivalent to assuming that all DMs are able to completely express their prior weights across the space of explanations and models and that each DM shares the same weight assignments as the analyst.

Uncertainty analyses generally expand, rather than restrict, the range of plausible $LR$ values. That is, values within a presented uncertainty range are necessarily plausible so long as the corresponding assumptions are considered reasonable; however, values outside a presented range are not necessarily implausible. Demonstrating a particular $LR$ value to be implausible requires first identifying the collection of all reasonable assumption sets and ensuring that no model within this collection gives rise to that value. However, the set of all reasonable models is nebulous, personal, and not optimizable in a normative sense. To begin to explore the relationship between data, assumptions and interpretations we consider multiple assumption sets in a form we refer to as the *lattice of assumptions*. We present the resulting ranges of $LR$s as an *uncertainty pyramid*. These are illustrated in a later section.

Without explicit statements of uncertainty, the lay audience may have a tendency to regard a given quantitative interpretation with undue mystique, reverence, and authority. This is one of several concerns expressed by Tribe (1971) in his seminal article titled "Trial by Mathematics: Precision and Ritual in the Legal Process" where he writes *"the very mystery that surrounds mathematical arguments – the relative obscurity that makes them at once impenetrable by the layman and impressive to him – creates a continuing risk that he will give such arguments a credence they may not deserve and a weight they cannot logically claim."*

When a statistical algorithm has been used to compute a $LR$, one may reasonably assume that the output of the algorithm, given its inputs, is highly reproducible and therefore has no substantial uncertainty. In this sense, a $LR$ value may be transferable as a discrimination score rather than as the ratio between posterior and prior odds. When viewed as a discrimination score, the $LR$ value does not abide probabilistic interpretation via Bayes equation, and its meaning is no longer self-contained. As with any score, its meaning only becomes apparent in the context of other $LR$ determinations, evaluated by the same process, in suitable, controlled reference applications. This information could be effectively conveyed for any method of assessing evidence through objective descriptions of protocols followed and outcomes obtained.

Even when it does not lead to a transferable measurement, the mathematical foundation described by Lindley provides useful guidance regarding questions to consider when evaluating the meaning of evidence.[9] Taking the DMs on a guided tour consisting of objective descriptions of demonstrable outcomes likely relevant to a particular investigation/analysis, including reference studies or other similar cases, provides the DMs an opportunity to answer these questions based on an accurate understanding of metrologically sound information.

---

[9]For example:
- How does the presented information demonstrate the frequency of correspondence that would occur between the crime scene evidence and realizations produced by the defendant obtained in accordance with the prosecution's explanation?
- What other explanations will be considered that do not imply the defendant's guilt?
- How does the presented information demonstrate the comparative infrequency of correspondence that would occur between the crime scene evidence and realizations produced in accordance with each of these explanations?

## Likelihood Ratio as the Pathway to Posterior Odds

We now illustrate the process of evaluating a $LR$ using more concrete notation. Suppose that evidence $y$ has been recovered from a crime scene and that, for simplicity, the only uncertain component across the explanations the DM is willing to consider for how $y$ came to be is the identity of its source.[10] Let $S_0, S_1, \ldots, S_N$ denote the totality of potential sources, one of which is responsible for $y$. The actual source of $y$ is denoted by $S_q$, where $q$ is unknown. The source $S_0$ is attributed to the defendant, and is of particular interest in that a DM will vote guilty if and only if he/she is convinced $S_0$ is the source of $y$. The event $E_0$, the truth of which is in doubt, is the following:

$$E_0 : S_0 \text{ is the source of } y \ (i.e., q = 0).$$

The complement $E_0^c$ of the event $E_0$ is

$$E_0^c : S_1 \text{ or } S_2 \text{ or } \ldots S_N \text{ is the source of } y \ (i.e., q \in \{1, 2, \ldots, N\}).$$

In addition to $y$, suppose one or more control samples (that is, samples from known sources) are available from one or more of the sources $S_j$, $j = 0, \ldots, N$. Denote these, collectively, by $x$.

Suppose $I$ denotes the totality of information available to the DM prior to being exposed to the information supplied by $y$ and $x$. According to the Lindley framework, a DM has prior probability $\pi_0 = Pr[E_0|I]$ for the event $E_0$ based on whatever information $I$ is available to him/her disjoint from $y$ and $x$. After being informed about the available new information $y$ and $x$, the DM would like to update his/her belief concerning the event $E_0$ in a *rational* and *coherent* manner.

The DM is interested in $Pr[E_0|y, x, I]$, the probability that $S_0$ is the source of $y$ given all the information available in the crime scene evidence ($y$), the control samples ($x$) and whatever else ($I$). Using the odds form of the Bayes rule, and following Lindley (1977), Neumann (2012, JRSS A) and others, we get

$$LR = \frac{Pr[y|x, E_0, I]}{Pr[y|x, E_0^c, I]}. \qquad (2)$$

In the context of this example, there is only one explanation under which the defendant would be considered guilty. Hence, the $LR$ numerator requires only the conditional probability of $y$ given $x$ and $E_0$ (and $I$). Suppose this is denoted by $Pr[y|x, E_0]$.[11]

When the number of possible alternative sources is greater than one, evaluating the $LR$ denominator, which corresponds to explanations under which the defendant is not guilty, is more complex. The event $E_0^c$ only means that $S_0$ is not the source of $y$ and, as such, it does not say anything about which of $S_1, \ldots, S_N$ is in fact the source (for simplicity of discussion we are assuming that there is a single source). We can decompose $E_0^c$ as the union of the events $E_j$, $j = 1, \ldots, N$, where

$$E_j : S_j \text{ is the source of } y.$$

---

[10] *E.g.*, if $y$ were a fingerprint, suppose the only relevant component of uncertainty to the DM is which person, or more specifically which finger, left the impression; or if $y$ consisted of striation marks on a bullet fragment, suppose the DM is only concerned about identifying the gun from which the bullet was fired.

[11] For simplicity of presentation, we have dropped the term $I$ with the proviso that all probabilities mentioned are *conditional on $I$*. Furthermore, it is to be understood that expressions such as $Pr[y]$ (or $Pr[y|x]$) refer to marginal (or conditional) probabilities or probability densities depending on whether $y$ is treated as discrete or continuous.

Because the corresponding hypothesis of not guilty involves multiple explanations, computing the $LR$ denominator requires both a weight and conditional probabilities of $y$ given $x$ and $E_j$ for each $E_j$, $j = 1, \ldots, N$. Suppose $\pi_0, \pi_1, \ldots, \pi_N$ are the *prior* probabilities from the perspective of the DM associated with the events $E_0$, $E_1$, ..., $E_N$, respectively. Then the denominator of the $LR$ takes the form

$$Pr[y|x, E_0^c] = \sum_{j=1}^{N} w_j Pr[y|x, E_j],$$

where $Pr[y|x, E_j]$ is the probability of $y$ given $x$ and $E_j$ (and $I$) and $w_j = \dfrac{\pi_j}{1 - \pi_0}$. Thus, $w_j$ are the prior probabilities of the DM associated with $E_1, \ldots, E_N$, given $E_0$ is false.

Given the quantities $Pr[y|x, E_j], j = 0, 1, \ldots, N$, and $\pi_0, \pi_1, \ldots, \pi_N$, the corresponding $LR$ is given by

$$LR = \frac{Pr[y|x, E_0]}{\sum_{j=1}^{N} w_j Pr[y|x, E_j]}.$$

Based on the rationale presented earlier, the assessment of prior odds is left to the DM and the forensic expert is tasked with the assessment of $LR$. Provided a $LR$ value by the forensic expert, a DM can obtain his/her corresponding posterior probability for $E_0$.[12] At least, that is the theory.

### List of Concerns

In this subsection, we identify issues requiring careful consideration before implementing the adapted Lindley framework favored by many forensic statisticians. Incomplete consideration of these issues and any steps taken to address them stands to jeopardize the judicial process.

1. **Whose Relevant Population?**
   According to the definition of the $LR$, any source given a prior probability greater than zero by the DM can influence the value of the $LR$ denominator and is therefore relevant; sources given prior probability zero cannot influence the value of the $LR$ denominator regardless of the value of the corresponding likelihood $Pr[y|x, E_j]$, and are therefore irrelevant to the DM. This set of sources with prior probability greater than zero forms the *relevant alternative population* for the DM. This is often simply referred to as the *relevant population*. If there are several DMs then each one could have their own set of weights $w_j$ and hence their own relevant population. Given a particular relevant population, the weights assigned to elements of that population can affect the $LR$ unless the assigned likelihoods are constant across all members of the population. In particular, $w_j = \dfrac{1}{N}$ is a special case, not a mandate.

2. **Whose Likelihoods?**
   In practice, probability functions $Pr[y|x, E_j], j = 0, 1, \ldots, N$ are rarely shared by all stakeholders. A common strategy is for the forensic analyst permit a distribution over a class of models, that will then be reweighted by consideration of empirical data. That is to assume, conditional on a parameter $\theta$ (typically finite dimensional, but possibly infinite dimensional) and the event $E_j$, crime scene data $y$ and control data $x$ are independently distributed with known distributions $g(y|\theta, E_j)$ and $h(x|\theta, E_j)$, respectively. It is further assumed that, given $E_j, j = 0, 1, \ldots, N$, $\theta$ has a known distribution described by the probability function $f(\theta|E_j)$.

---

[12] As discussed earlier, though Lindley (2014) makes it clear that the $LR$ in Bayes formula is the personal $LR$ of the DM, many treat $LR$ as a quantity that can be (and should be) calculated by a forensic expert; see, for instance, (Aitken and Taroni (2004), chapter 3).

The function $f(\theta|E_j)$ is used to express a prior belief about likelihood functions for $x$ and $y$ given $E_j$ (not to be confused with the prior $\pi_j$, which reflects prior belief in the event $E_j$). Hence the joint distribution of $y, x$, and $\theta$, given the event $E_j$, is described by the probability function

$$a(y, x, \theta|E_j) = g(y|\theta, E_j)h(x|\theta, E_j)f(\theta|E_j). \tag{3}$$

It is now an exercise in calculus to obtain the quantity $Pr[y|x, E_j]$ as

$$Pr[y|x, E_j] = \frac{\int a(y, x, \theta|E_j)d\theta}{\int \int a(y, x, \theta|E_j)d\theta dy} \tag{4}$$

Note that the joint distribution of all relevant quantities has been assumed to be *exactly* known through the choice of $f$, $g$ and $h$. The distribution of interest for source $j$, $Pr[y|x, E_j]$, is a conditional distribution that follows from the assumed joint distribution $a(y, x, \theta|E_j)$. Asymptotically, as the number of control observations goes to infinity for each potential source $j = 0, 1, \ldots, N$, the value of $Pr[y|x, E_j]$ will converge to the same answer for many different choices of $f$, $g$ and $h$. With finite data, however, choices of $f$, $g$ and $h$ remain important sources of uncertainty across stakeholders.

Occasionally, support for particular choices of $f$, $g$ and $h$ is given by showing them to be consistent (as defined by some user-selected process for evaluating such things) with empirical data from similar situations. The merits of these presentations often depend on the available data being representative of other results one might obtain in a different or more extensive data collection effort. Statistical uncertainty corresponding to such generalizations is calculable only when the observed sample was collected according to a probability sampling method.[13] In the absence of a well-defined population and a corresponding probability sampling method, the representativeness of data is a matter of personal belief, generally based on the inability to identify any particular factor that would cause the sample to be unrepresentative. Probability sampling methods are designed to ensure that no factor has such an effect, regardless of whether or not it is identified.

Even when representative data are used to guide modeling choices, the existence of alternative choices that are consistent with any available empirical guiding information is generally overlooked, and the impact that making alternative choices has on the end quantity is rarely, if ever, extensively explored. The likelihoods $Pr[y|x, E_j]$, and the resulting $LR$, may be greatly influenced by the choices made by the forensic expert alone. One way to overcome this deficiency is to carry out a more complete exploration of the effects the choices have on the reported $LR$ using what we call the Lattice of Assumptions. We will illustrate this concept in a later section in the context of an example.

3. **Incoherence**

The framework proposed by Lindley (2014, chapter 10, page 260) and backed by the arguments of Bayesian decision theory lead to the equations

$$\text{Posterior Odds}_{\text{DM}} = \text{Prior Odds}_{\text{DM}} \times LR_{\text{DM}}$$

---

[13]For finite populations, probability sampling methods refer to strategies for selecting a subset of a collection in a manner such that each element's probability of being selected is greater than zero and the selection probability for each subset is known. In the case of infinite populations, sampling generally requires discretization to facilitate implementation.

from the perspective of the DM and

$$\text{Posterior Odds}_{\text{Expert}} = \text{Prior Odds}_{\text{Expert}} \times LR_{\text{Expert}}$$

from the perspective of the expert. As pointed out earlier, the proposed adaptation is a hybrid of these perspectives represented as

$$\text{Posterior Odds}_{\text{DM}} = \text{Prior Odds}_{\text{DM}} \times LR_{\text{Expert}}$$

which is no longer supported by the rationality and coherence arguments of Bayesian decision theory and leaves only a single degree of freedom to the DM, viz., the assessment of $\pi_0 = Pr[E_0]$. Genuinely applying Bayesian decision theory upon consideration of whatever has been communicated by the expert, say $T$, would require that each DM compute his/her own $LR$ for $T$,

$$LR_{\text{DM}}(T) = \frac{Pr_{\text{DM}}(T|H_p)}{Pr_{\text{DM}}(T|H_d)}.$$

The risk of misunderstanding and misapplication seems high when $T$ consists of the personal $LR$ of an expert with whom DMs are unlikely to have any personal history or interaction. *This risk would be further increased following misdirection as to how DMs are to use the provided number to obtain their own posterior probabilities.*

4. **Restrictive Application of Decision Theory**
   Under Bayesian decision theory, the DM is considered *rational* when they reach a decision by maximizing their expected utility, $EU$. That is, for each potential act, the DM calculates the weighted sum of utilities

$$EU(Act_i) = \sum Utility(Act_i, Explanation_j) \times Pr(Explanation_j | Evidence) \qquad (5)$$

and picks the act for which $EU$ is largest. In essence, DMs should decide to act in the manner that makes them happiest on average, according to their own beliefs and preferences.

When posterior probabilities are assigned to groups of explanations, rather than each of the individual explanations, expected utility can be maximized only when the utility functions for any two explanations $j$ and $k$ in a common group satisfy the constraint

$$Utility(Outcome_{ij}) = Utility(Outcome_{ik}) + c_{jk}$$

for all $i$, where $c_{jk}$ is a scalar and $Outcome_{ij}$ is the result occurring if explanation $k$ is exclusively true and DM chooses act $i$.

In theory, there is no constraint as to how many groups are required to satisfy this constraint; maximizing expected utility may require posterior probabilities for each of many finely resolved sets. The proposed usage of a $LR$ requires that the explanations be partitioned into two groups, and only facilitates posterior probability determination at this coarse scale. The point here is that even if DMs decide to adopt an expert's $LR$ as their own, they may still not be able to apply Bayesian decision theory.

5. **Restrictive Meaning of Probability**
   When following a subjective Bayesian approach one uses a definition of *personal* probability that could be viewed as an individual's assessment of a fair value for a bet of $E_0$ versus its complement. It is assumed that there is a unique such value and that the individual is able to identify this value without any doubts.[14] Moreover, it is assumed that this individual (DM) is able to specify all needed personal probabilities in this manner and the collection of specified probabilities satisfy the requirement of *coherence.* Lindley *et al.*, (1979) discuss the practical issues one must address in order to reconcile the generally incoherent probability assessments by an individual. They consider several different approaches that one could use in such a reconciliation process. See also Kadane and Winkler (1985). The fact that such reconciliation efforts are necessary points to uncertainties associated with subjective probability assessments. Nevertheless, results derived using such probability models are treated as free from uncertainties (see, for instance, Taroni, *et al.* 2015).

## The Restrictive Influence of Modeling Assumptions

We are concerned about seemingly innocuous modeling assumptions latently constraining the space of plausible interpretations as might be presented by a forensic expert. In this section we demonstrate a process for evaluating the restrictive influence of unsubstantiated information that can creep in solely on the basis of distributional assumptions made by an analyst. It should be noted that the data and modeling approaches used in this section are not exhaustive and are not intended to represent analyses generally undertaken by any particular forensic practice. As such, the actual numerical results obtained in this section are not of primary interest. Our intention is to illustrate an application of assessing the influence of modeling assumptions on a concrete example.

To evaluate the influence of a given assumption, one should consider the results of several analyses, one in which the assumption was made (say, assumption $A$) and others in which different assumptions (say, assumption $B_i, i = 1, 2, \ldots$), consistent with empirically observed data, are made. The influence of the assumption is reflected by the differences between the conclusions drawn upon evaluation of each set of results. In cases where the differences are considered to be substantial, assumption $A$ has played a critical role, and the conclusion reached from results of the analysis in which assumption $A$ was made stretches beyond what the data itself, used in the analysis, can support. In such a case, the basis of the appropriateness of any particular assumption among the collection $A$ and $B_i, i = 1, 2, \ldots$ should be carefully considered. To aid in this effort, the investigator should describe, and display where possible, the basis of their perception of the extent to which the assumption may be inaccurate.

### Illustration: Glass Example

We consider an educational example discussed in chapter 10 of Aitken and Taroni (2004) involving measurements of refractive indices (RI) for glass evidence. Suppose a window is broken during the commission of a crime and fragments of glass are recovered from the crime scene. Suppose also that fragments of glass were found on a suspect. Denote by $x_1, \ldots, x_m$ the RIs of the crime scene fragments (bulk sample) and by $y_1, \ldots, y_n$ the RIs of suspect fragments (receptor sample). The two propositions of interest are

$H_p$ : The receptor sample is from the same source as the bulk sample

$H_d$ : The receptor sample is from a different source than the bulk sample.

---

[14] For a systematic introduction to *imprecise probabilities* see, for instance, Walley (1991). This field remains an active area of research (Augustin, *et al.,* 2016)

In this illustration we are operating as though there is no information available for us to justify assigning different likelihoods to each particular potential source. Hence, we consider the probabilities in the numerator and the denominator of the LR from the perspective of a population of windows rather than weighting likelihoods from individual windows according to their prior probability; see equation (6) and comments following equation (1).

**Within Source and Between Sources Distributions**

Interpreting the information contained in the observed RIs regarding these two propositions requires understanding the distribution of RIs within each source and how that distribution varies from one source to the next. (Note that if the RI distribution did not vary between sources, then the RI observations would not provide any useful information about their source.) Consideration of how the RI distribution varies from one glass pane to the next results in a distribution of distributions. The collection of possible descriptions or models for the distribution of distributions is overwhelmingly vast. The tendency is to limit the class of potential descriptions by specifying properties of RI distributions that are assumed to remain constant from one window to the next. In particular, the RI distributions across glass panes are often assumed to be identical except for their location (*e.g.*, mean or median). That is, the RI distribution for every potential source is assumed to have exactly the same shape and exactly the same scale (or spread). This assumption implies that the collection of values obtained by subtracting, from RIs of each fragment within the glass pane, the median of RIs of all fragments from the glass pane (*i.e.*, $x - \text{median}(x)$), would have exactly the same distribution for any glass pane in the considered relevant population.

In general, the results of analysis (*e.g.*, $LR$) can be highly sensitive to deviations from the assumption that RI distributions differ only by their median from one glass pane to another. Generating empirical confidence in such a strong assumption would require collecting RI data from many windows with enough measurements from each window so as to convince oneself that strictly limiting the set of plausible distributions to a location family will only have a *negligible effect* on the interpretation of the analysis results compared to, for instance, when the shape and scale of the presumed location family are allowed to vary from one source to another. Even with such a vast and consistent dataset, the possibility remains that the RI distribution of any unexamined window differs substantially from the observed characteristics of the other windows. Further illustration of the potential influence of assuming a location family on the interpretation of the observed RI from a particular case is beyond the scope of this paper. That is, the notion of uncertainty we portray in these examples is incomplete. The uncertainty resulting from a more complete examination is expected to be greater than what is illustrated here.

For the sake of simplicity, we proceed by supposing that the informed DM is willing to make the location family assumption. To compute a $LR$ for this scenario, let us first introduce some notation. Suppose the cumulative distribution function ($CDF$) of RI values from any single window belongs to the location family of distributions $G(y;\theta) = G_0(y - \theta)$ for some continuous distribution with $CDF$ $G_0$ whose median value is zero. Denote the corresponding probability density function ($PDF$) by $g_0$. Furthermore, suppose that, across the (relevant) population of windows, the median RIs $\theta^{(j)}, j = 0, 1, \ldots, N$ are iid with (an unknown) density function $f(\theta)$ and corresponding $CDF$ $F(\theta)$. That is, we have assumed that $f(\theta|E_j) = f(\theta)$ for all $j = 0, 1, \ldots, N$. For completeness, we display the expression for the resulting $LR$ in

Equation (6).

$$LR = \frac{\int \left(\prod_{i=1}^{m} g_0(x_i - \theta)\right) \left(\prod_{j=1}^{n} g_0(y_j - \theta)\right) dF(\theta)}{\left(\int \left(\prod_{i=1}^{m} g_0(x_i - \theta)\right) dF(\theta)\right) \left(\int \left(\prod_{j=1}^{n} g_0(y_j - \theta)\right) dF(\theta)\right)}. \quad (6)$$

**Aitken and Taroni Illustrative Analyses**

In the illustrative example provided in Aitken and Taroni (2004), it is assumed that $g_0$ is the $PDF$ of a normal distribution with a standard deviation equal to 0.00004. That is, the collection of RIs that could be observed from windows are independently and identically distributed (iid) according to a normal distribution with unknown window-specific mean $\theta^{(j)}, j = 1, \ldots, N$, and known standard deviation $\sigma$ equal to 0.00004. Lambert and Evett (1984), in their Table 10.5, give average RI measurements from 2269 different samples of float glass. Assuming that these measurements are representative of the mean RIs associated with sources $S_j, j = 0, 1, \ldots, N$, Aitken and Taroni apply kernel density estimation, using a Gaussian kernel, with varying bandwidths to estimate the density $f$ (or the $CDF$ $F$) from these sample data. The resulting estimates are then used to evaluate the $LR$ corresponding to various hypothetical pairs of average RI measurements from the source (window) and receptor (suspect) (see Table 10.6, page 341, Aitken and Taroni 2004). Applying the distribution estimates from Aitken and Taroni to the illustrative example from Evett (1977) (see Table 1), accounting for interval censoring of the recorded measurements to plus or minus 0.00001, produced corresponding $LR$s of 196, 184, and 72, respectively.

**Table 1:** Refractive Index Measurements from the window and from the suspect

| Measurements from the window | 1.51844 | 1.51848 | 1.51844 | 1.51850 | 1.51840 |
|---|---|---|---|---|---|
| Measurements from the suspect | 1.51848 1.51848 | 1.51846 1.51850 | 1.51846 1.51848 | 1.51844 1.51844 | 1.51848 1.51846 |

**Multiple Plausible Models**

The consideration of multiple kernel bandwidths for estimating $f$ begins to illustrate the potential uncertainty due to the influence of modeling choices. A more complete evaluation may be obtained by considering the set of all combinations of $g_0$ and $f$ that might be considered plausible and how variable the computed $LR$ is across combinations within that set. Plausibility is not an objective notion. The criteria for establishing the plausibility of a given proposal is likely to vary from one person to the next. However, it is possible for the criteria of a specific individual to be expressed in an objective manner. When criteria for plausibility have been es-tablished[15], it is more complete to consider the set of all plausible models (possibly obtained by filtering the class of all models through the plausibility criteria) rather than identifying a single plausible model (or a narrow set of closely related models in the case of multiple estimates of $f$ obtained from different bandwidths) and proceeding as though it is the only plausible model or representative of all plausible models.

**Goodness-Of-Fit Tests and Plausibility Criteria**

We note that it is common practice for a data analyst to use a statistical test of goodness-of-fit to assess plausibility of one or more models. In our example, the data modeler could assess the plausibility of a proposed distribution pair ($g_0$ and $f$), given sample data, using any of a number of goodness-of-fit statistical testing procedures. Some well-known methods are: (1) Kolmogorov-Smirnov (KS) test, (2) Cramer-von Mises (CvM) test, (3) Anderson-Darling (AD) test. For related other approaches the interested reader should also consult Owen (1995), Goldman and Kaplan (2015), Liu and Tewfik (2013) and Frey (2008).

Here we consider the KS test for illustrative purposes. Any other procedure can be used in place of the KS test but the computations can be more challenging. The concept is the same for each criterion: the sample data itself cannot reduce the space of plausible models to a single $CDF$. The KS test leads to a confidence band for $CDF$s that are consistent with the data at a prescribed level of confidence, say 95%. When the KS test is used to assess plausibility, any $CDF$ that lies entirely within the confidence band would be deemed plausible given the sample data. As the number of observations in the data set increases, the confidence band narrows and the set of plausible distributions is reduced.

**Between-Windows and Within-Window Data Sets for the Glass Example**

We now consider the influence of two data sets on plausible choices for $g_0$ and $f$.

**Float Glass Data**

The first data set (See Lambert and Evett, 1984) contains a collection of average RI measurements obtained from various within-window samples collected from different manufactured pieces of float glass. The number of observations contained in each sample is not provided, so sample sizes may vary across the samples and there is some uncertainty as to how this data should be viewed during evidence evaluation. If each sample contained a single observation, the KS confidence band might be used to restrict the marginal distribution of a single RI measurement obtained from a randomly selected window in the population. This marginal distribution is determined by the choice of $g_0$ and $f$ as $h(y) = \int g_0(y - \theta) dF(\theta)$. If the samples consisted only of means of many replicate observations, the KS bounds could serve to restrict the class of plausible choices for $f$, but would not provide much insight to the choice of $g_0$.

For illustrative purposes, we treat the data from this set as providing median RI values for a sample of 2269 windows representative of the relevant population. We use the median rather than the mean to reduce the sensitivity of the location parameter $\theta$ to the tails of the distribution $g_0$, which cannot be well-estimated from sample data. The 2269 reported

---

[15] Analogous to selecting prior distributions when conducting Bayesian inference, the choice of a plausibility criterion should not be guided by the set of $LR$ values they permit, but upon information available before application to the case at hand. Additionally, while here we focus on the application of a plausibility criterion directly to the space of modeling choices, the same concept applies to prior specification in a Bayesian inference framework. That is, one specifies the criteria by which a prior is determined to be acceptable and seeks to filter the space of all possible priors through the plausibility criterion.

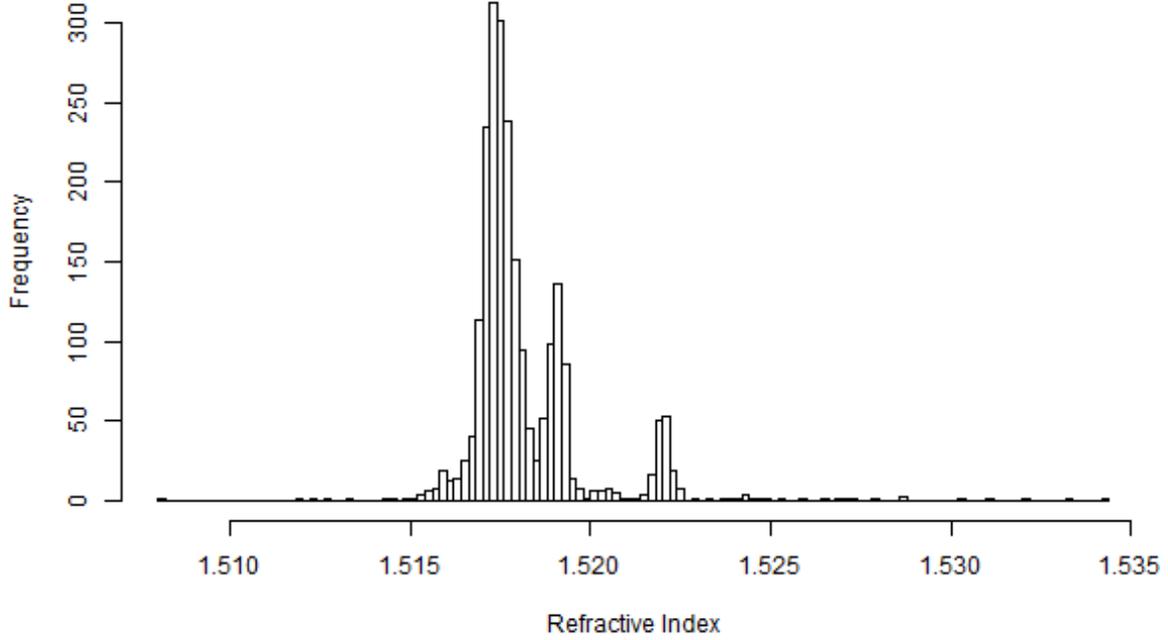

**Figure 1:** Histogram of Glass Data

RI values are shown in Table 2. A histogram of these data is shown in Figure 1. Figure 2 shows the empirical $CDF$ ($eCDF$) for these data along with the lower and upper boundaries of a KS 95% confidence band used to define which choices for $f$ will be considered plausible given the $eCDF$. In the lattice of assumptions illustration, we consider several estimates for $f$ based on Gaussian kernel density estimates fit to the 2269 observations with bandwidths spanning from 0 (which corresponds to the $eCDF$) to $2.155 \times 10^{-4}$, which is the maximum bandwidth for which the corresponding discrete distribution obtained by accounting for the reported measurements being interval censored (to plus or minus $1 \times 10^{-4}$) remains entirely within the KS confidence band. $CDF$s for the discrete distributions obtained by accounting for interval censored measurements and the corresponding underlying continuous distributions are shown in Figure 2 for both the $eCDF$ and the smoothest kernel density estimate. Kernel density estimates resulting from the intermediate bandwidths of $10^{-5}$, $2 \times 10^{-5}$, $5 \times 10^{-5}$, and $10^{-4}$ are considered during computation but are not displayed. For illustration only, we also include a $CDF$ not produced by kernel density estimation. This $CDF$, referred to as Jump, follows the lower KS bound for values less than the mean RI value $m = \dfrac{\sum_{i=1}^{10} y_i + \sum_{j=1}^{5} x_j}{15}$ for the 15 sample fragments, and the upper KS bound for values greater than $m$, with a jump at $m$. This $CDF$ is shown in blue in Figure 2.

**Table 2:** Refractive index measurements for 2269 glass fragments given in Lambert and Evett (1984).

| RI | Count | RI | Count | RI | Count | RI | Count |
|---|---|---|---|---|---|---|---|
| 1.5081 | 1 | 1.5170 | 65 | 1.5197 | 7 | 1.5230 | 1 |
| 1.5119 | 1 | 1.5171 | 93 | 1.5198 | 1 | 1.5233 | 1 |
| 1.5124 | 1 | 1.5172 | 142 | 1.5199 | 2 | 1.5234 | 1 |
| 1.5128 | 1 | 1.5173 | 145 | 1.5201 | 4 | 1.5237 | 1 |
| 1.5134 | 1 | 1.5174 | 167 | 1.5202 | 2 | 1.5240 | 1 |
| 1.5143 | 1 | 1.5175 | 173 | 1.5203 | 4 | 1.5241 | 1 |
| 1.5146 | 1 | 1.5176 | 128 | 1.5204 | 2 | 1.5242 | 1 |
| 1.5149 | 1 | 1.5177 | 127 | 1.5205 | 3 | 1.5243 | 3 |
| 1.5151 | 1 | 1.5178 | 111 | 1.5206 | 5 | 1.5244 | 1 |
| 1.5152 | 1 | 1.5179 | 81 | 1.5207 | 2 | 1.5246 | 2 |
| 1.5153 | 1 | 1.5180 | 70 | 1.5208 | 3 | 1.5247 | 2 |
| 1.5154 | 3 | 1.5181 | 55 | 1.5209 | 2 | 1.5249 | 1 |
| 1.5155 | 5 | 1.5182 | 40 | 1.5211 | 1 | 1.5250 | 1 |
| 1.5156 | 2 | 1.5183 | 28 | 1.5212 | 1 | 1.5254 | 1 |
| 1.5157 | 1 | 1.5184 | 18 | 1.5213 | 1 | 1.5259 | 1 |
| 1.5158 | 7 | 1.5185 | 15 | 1.5215 | 1 | 1.5265 | 1 |
| 1.5159 | 13 | 1.5186 | 11 | 1.5216 | 3 | 1.5269 | 1 |
| 1.5160 | 6 | 1.5187 | 19 | 1.5217 | 4 | 1.5272 | 2 |
| 1.5161 | 6 | 1.5188 | 33 | 1.5218 | 12 | 1.5274 | 1 |
| 1.5162 | 7 | 1.5189 | 47 | 1.5219 | 21 | 1.5280 | 1 |
| 1.5163 | 6 | 1.5190 | 51 | 1.5220 | 30 | 1.5287 | 2 |
| 1.5164 | 8 | 1.5191 | 64 | 1.5221 | 25 | 1.5288 | 1 |
| 1.5165 | 9 | 1.5192 | 72 | 1.5222 | 28 | 1.5303 | 2 |
| 1.5166 | 16 | 1.5193 | 56 | 1.5223 | 13 | 1.5312 | 1 |
| 1.5167 | 15 | 1.5194 | 30 | 1.5224 | 6 | 1.5322 | 1 |
| 1.5168 | 25 | 1.5195 | 11 | 1.5225 | 3 | 1.5333 | 1 |
| 1.5169 | 49 | 1.5196 | 3 | 1.5226 | 5 | 1.5343 | 1 |

**Bennett Data**

The second data set consists of 49 refractive index measurements on samples of fragments from 49 different locations on a single window and is used to evaluate the plausibility of within-window distribution choices. These data were collected by Bennett et. al (2003) and are also mentioned in Curran (2011) (see page 42).[16] They are publicly available in the dafs package in R. The original data set consists of RI measurements for a sample of 10 fragments from each of 49 locations on a single window pane for a total of 490 readings. We have selected a single fragment from each of the 49 locations (the listed value in the first row of the bennett.df data frame in dafs). These data are reproduced in Table 3 for the convenience of the reader. For illustrative purposes we treat these 49 RI values as representative of the RI distribution within a single window, providing guidance for choosing $g_0$.

For the 49 RI measurements in Table 3, the empirical $CDF$ and corresponding KS 95% confidence band are shown in Figure 3. In the lattice of assumptions we consider several distributional shapes including those pertaining to a normal distribution, $t$-distributions with 1 and 0.14 degrees of freedom, respectively, and $\chi^2$ distributions with 2 and 3 degrees of freedom. We also consider a small simulated collection of *nonparametric CDF*s (not belonging to any particular parametric family). Some of these nonparametric $CDF$s fulfill additional constraints of unimodality and/or symmetry. For each considered distributional shape, we identify the range of scale parameters such that the discrete distribution obtained by accounting for the interval-censoring of the reported within-window measurements is contained entirely within the confidence band. For each shape, we consider estimates of $g_0$ obtained at 15 evenly spaced scale values spanning this range. For a given shape and scale parameter, the $LR$ is evaluated for each pairing of $g_0$ with each of the choices for $f$ described above. Figure 3 provides a visual summary of the analysis when the shape of $g_0$ is assumed to be that of a normal distribution. Analogous displays for other considered shapes are provided in the Appendix.

---

[16]Although not explicitly mentioned in Bennett et. al (2003), these data appear to be interval-censored with variable interval half-widths (approximately) equal to $1.5 \times 10^{-6}$. Consequently, all of our analyses based on these data take this interval-censoring into account

**Table 3:** Refractive Index Measurements from 49 different locations from a single window. (Data courtesy of Curran (2011))

| | | | | | | |
|---|---|---|---|---|---|---|
| 1.519788 | 1.519901 | 1.519941 | 1.519941 | 1.519941 | 1.519963 | 1.519970 |
| 1.519974 | 1.519974 | 1.519974 | 1.519974 | 1.519974 | 1.519978 | 1.519978 |
| 1.519978 | 1.519981 | 1.519981 | 1.519981 | 1.519981 | 1.519985 | 1.519989 |
| 1.519989 | 1.519992 | 1.519992 | 1.519996 | 1.519996 | 1.519996 | 1.519996 |
| 1.520000 | 1.520000 | 1.520003 | 1.520007 | 1.520007 | 1.520007 | 1.520007 |
| 1.520010 | 1.520010 | 1.520014 | 1.520014 | 1.520014 | 1.520014 | 1.520025 |
| 1.520025 | 1.520029 | 1.520040 | 1.520043 | 1.520047 | 1.520047 | 1.520069 |

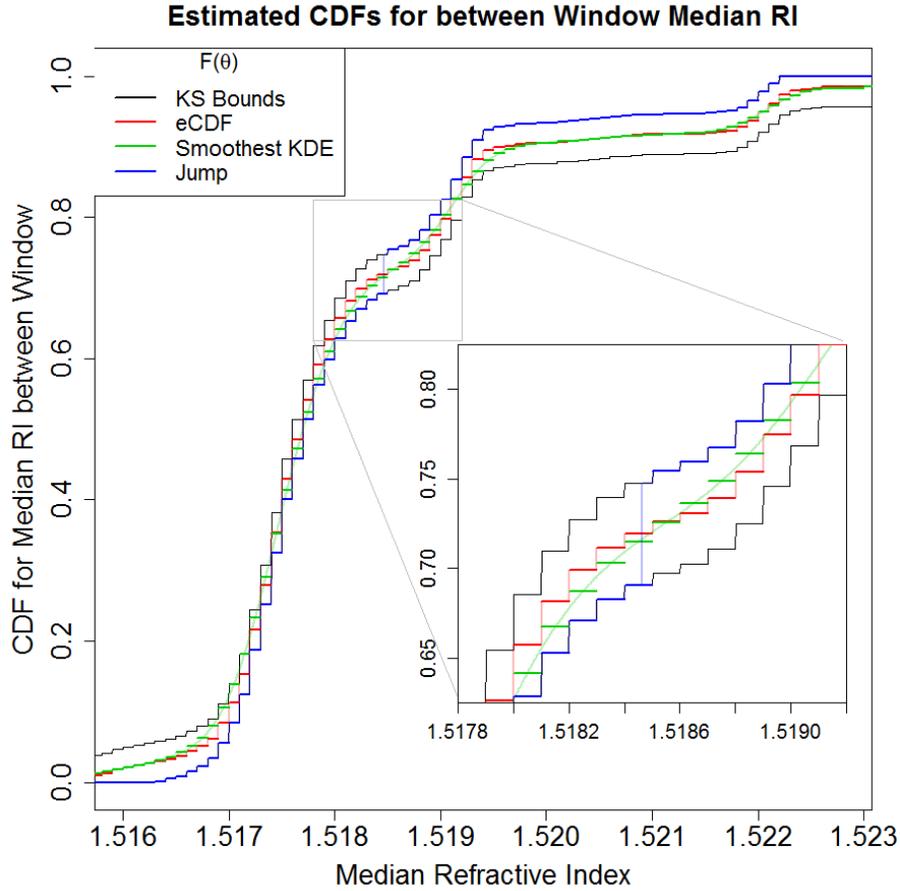

**Figure 2:** 95% Kolmogorov-Smirnov Confidence Band for the Lambert and Evett Glass Data. The bold line segments portray the discrete distribution obtained by accounting for the reported data being interval censored to $\pm\, 0.0001$. The faded lines display the $CDF$ of the underlying continuous distribution.

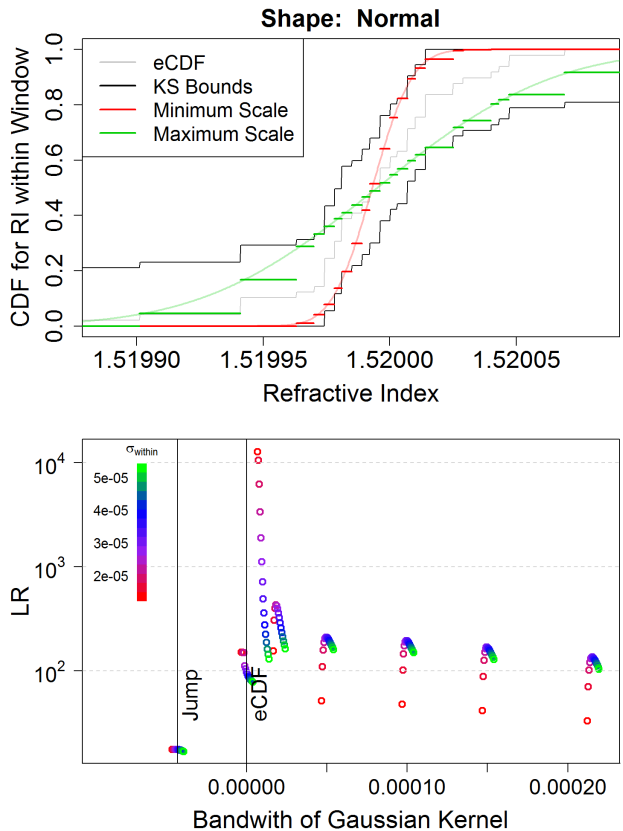

**Figure 3:** Top: 95% Kolmogorov-Smirnov confidence band for the $CDF$ of refractive indices from 49 fragments from a single window (Bennett Data). The empirical $CDF$ is shown in gray. The faded red and green smooth curves respectively correspond to normal distributions with the smallest and largest scale (standard deviation) parameters such that the discrete distributions obtained, to account for interval-censoring in the reported data (shown using solid red and solid green line segments, respectively), are entirely contained within the confidence band.
Bottom: $LR$ values corresponding to various choices of $F$, reflected by position along the x-axis, and the scale factor for the shape corresponding to a normal distribution. The left-most results correspond to the estimate of $F$ labeled as Jump, which is displayed in Figure 2. The remaining positions reflect the bandwidth of the Gaussian kernel leading to the estimate of $F$ used in computing the $LR$. Within each choice of $F$, the $LR$ values are staggered in order of the scale parameter used to define $g_0$ to emphasize the potential non-monotonic relationship between $LR$ and scale parameter. The points are color coded to indicate the associated scale parameter values.

## Lattice of Assumptions

When modeling the distribution of RI values for fragments from any single window, Lindley (1979) assumed normality as do Aitken and Taroni (2004). We recognize this was done for illustrative purposes only. Nevertheless, it is worth noting that normal distributions represent a tiny fraction of $CDF$s meeting the KS criteria, and the impact of exclusively assuming a normal distribution is not clear until the set of $LR$ values obtainable by using other distributions lying within the KS bounds have been investigated. In recognition that a given individual's criterion for a distribution to be plausible may include conditions beyond a KS test, in this section we examine the $LR$ values obtainable by distributions satisfying a variety of assumption sets. These assumption sets are displayed in the form of a lattice diagram (Grätzer, 2011) as shown in Figure 4. In the figure, when a line segment connects two assumption statements, the assumption appearing lower on the lattice diagram is nested within (*i.e.*, more restrictive than) the assumption appearing higher. In Table 4 and Figure 5, we report interval summaries of the range of $LR$ values over the considered subset of the space of all possible models satisfying each node's criteria.

## Discussion of Results

Results in Table 4 and Figure 5 clearly demonstrate that within this particular educational example the distributional assumptions made regarding the data generating process can have a substantial effect on the $LR$ values that would be reported. Keep in mind that we have only examined a (small) subset of all possible $CDF$s that would be deemed by the KS confidence band to be consistent with the considered RI data. As such, the uncertainty pyramid portrayed in Figure 5 is likely to under-represent the influence of choices of $f$ and $g_0$ within this example. Once again, the point is that reporting a single $LR$ value after an examination of available forensic evidence fails to correctly communicate to the DM the information actually contained in the data. Personal choices strongly permeate every $LR$ value. If expert testimony is to include the computation of a $LR$, we feel an assumptions lattice and corresponding $LR$ uncertainty pyramid provide a more honest assessment of the information in the evidence itself.[17]

---

[17] The proposal to present an uncertainty pyramid is not intended to lessen the importance of providing objective descriptions of empirical results from analysis and investigation.

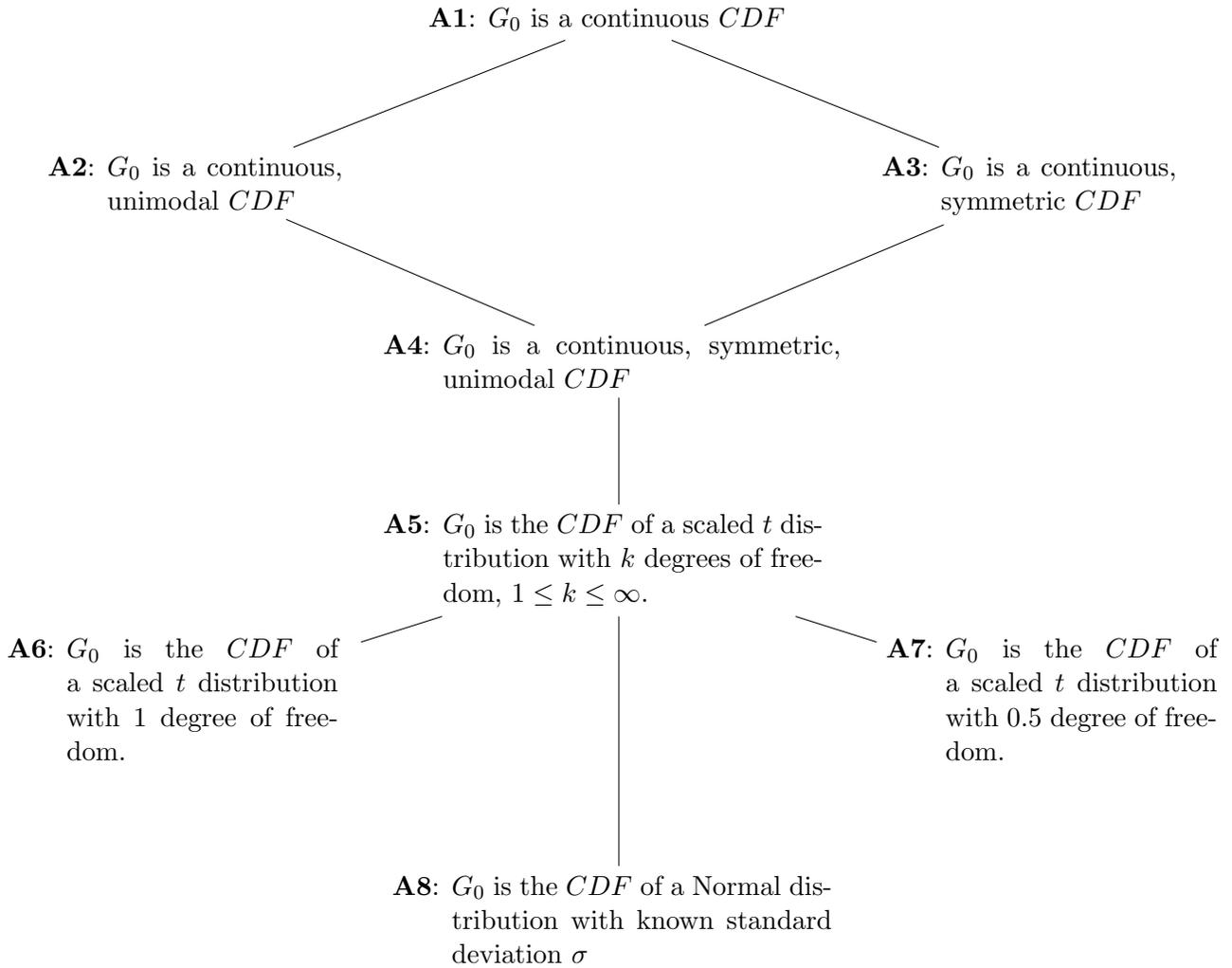

**Figure 4:** Assumptions Lattice for the Glass Example

**Table 4:** $LR$ values corresponding to different choices for $G_0$ and $F$.

| $G_0$ Shape | $G_0$ Scale | Bandwidth for $F(\theta)$ | $LR$ |
|---:|---:|---|---:|
| Normal | 1.24e-05 | (1e-05) | 1.28e+04 |
| Normal | 1.24e-05 | (0.00022) | 3.32e+01 |
| Normal | 1.24e-05 | Jump | 1.76e+01 |
| Normal | 5.45e-05 | Jump | 1.67e+01 |
| t with 1 d.f. | 7.31e-06 | (1e-05) | 5.20e+02 |
| t with 1 d.f. | 7.31e-06 | (0.00022) | 3.56e+01 |
| t with 1 d.f. | 1.69e-05 | Jump | 1.73e+01 |
| t with 1 d.f. | 4.10e-05 | Jump | 1.70e+01 |
| t with 0.14 d.f. | 1.15e-07 | (1e-05) | 1.03e+03 |
| t with 0.14 d.f. | 1.15e-07 | (5e-05) | 2.90e+01 |
| t with 0.14 d.f. | 4.03e-07 | Jump | 7.44e+00 |
| t with 0.14 d.f. | 1.15e-07 | Jump | 5.58e+00 |
| Non-parametric Symmetric Unimodal | 5.78e-01 | (1e-05) | 1.66e+04 |
| Non-parametric Symmetric Unimodal | 2.45e+00 | (0) | 1.54e+00 |
| Non-parametric Symmetric Unimodal | 3.07e+00 | Jump | 4.32e+01 |
| Non-parametric Symmetric Unimodal | 6.87e-01 | Jump | 2.96e-01 |
| Chi-squared with 3 d.f. | 6.37e-06 | (0) | 1.26e+02 |
| Chi-squared with 3 d.f. | 6.37e-06 | (5e-05) | 4.16e-08 |
| Chi-squared with 3 d.f. | 6.37e-06 | Jump | 1.26e+02 |
| Chi-squared with 3 d.f. | 1.06e-05 | Jump | 3.28e-09 |
| Chi-squared with 2 d.f. | 8.66e-06 | (0) | 1.26e+02 |
| Chi-squared with 2 d.f. | 8.66e-06 | (0.00022) | 1.76e-07 |
| Chi-squared with 2 d.f. | 8.66e-06 | Jump | 1.26e+02 |
| Chi-squared with 2 d.f. | 1.75e-05 | Jump | 2.63e-07 |
| Non-parametric Unimodal | 5.02e-01 | (1e-05) | 2.79e+05 |
| Non-parametric Unimodal | 1.02e+00 | (0) | 6.71e-02 |
| Non-parametric Unimodal | 3.12e-01 | Jump | 9.15e+02 |
| Non-parametric Unimodal | 5.22e-01 | Jump | 5.75e-02 |
| Non-parametric Symmetric | 1.10e+00 | (1e-05) | 7.17e+04 |
| Non-parametric Symmetric | 3.82e-01 | (0) | 1.61e-08 |
| Non-parametric Symmetric | 2.91e-01 | Jump | 9.15e+02 |
| Non-parametric Symmetric | 3.58e-01 | Jump | 7.74e-08 |
| Non-parametric | 9.13e-01 | (1e-05) | 1.64e+05 |
| Non-parametric | 1.11e+00 | (0) | 7.49e-12 |
| Non-parametric | 2.46e-01 | Jump | 9.15e+02 |
| Non-parametric | 5.14e-01 | Jump | 8.41e-12 |

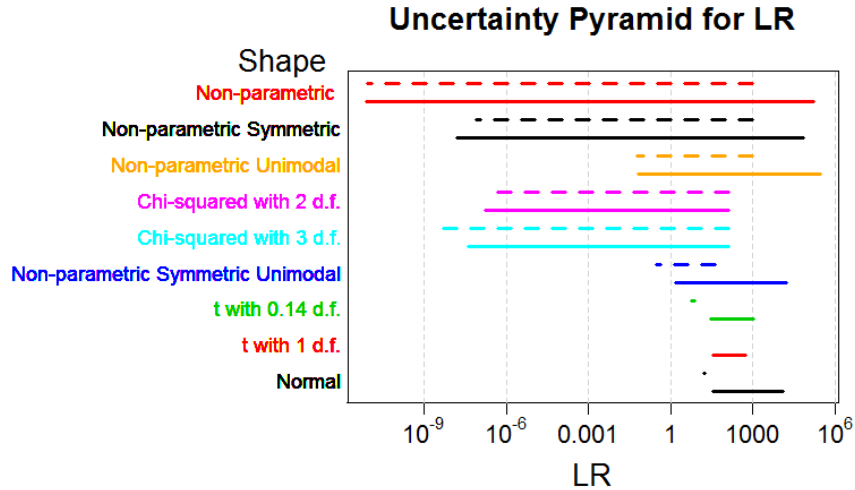

**Figure 5:** Ranges of $LR$ values corresponding to the choices of different assumptions for $F$ from the assumptions lattice combined with choices for $G_0$. Solid lines show the range of $LR$ values when $F$ is obtained by kernel density estimation. Dashed lines correspond to range of $LR$ values when using the Jump estimate of $F$.

## Summary

The $LR$ framework has been portrayed as having an exclusive, normative role in forensic expert communication on the basis of arguments centered around mathematical definitions of rationality and coherence (Biedermann *et al.*, 2015). These arguments are aimed at ensuring a form of self-consistency of a single, autonomous decision maker (DM).[18] However, decision theory does not consider the transfer of information between multiple parties as occurs throughout the judicial process when one or more DMs rely on forensic experts to help inform their decisions. Thus, while the normative role of decision theory may apply to a DM processing information presented during a case or trial in accordance with her own personal beliefs and preferences, it does not dictate the means by which a forensic expert should communicate the information to be considered.[19]

Bayesian decision theory neither mandates nor proves appropriate the acceptance of a subjective interpretation of another, regardless of training, expertise or common practice. Any normative properties of the process do not support any particular subjective inputs to the calculation of a $LR$ and therefore do not support any particular result. For anyone other than the analyst who computed it, a reported $LR$ value cannot be assumed to provide a direct path to obtaining a personal posterior probability. The suitability of a $LR$ value for use in Bayes formula depends on strictly personal choices and has the potential to vary substantially from one interpreter of the information to the next. This critical uncertainty component is not conveyed by the $LR$ value itself or by a list of assumptions behind it. The act of transferring information may be more constructively guided by considerations of metrology than of decision theory.

---

[18] More specifically, the Dutch book arguments (Hájek, 2008)

[19] Some may argue that because any given DM is likely unfamiliar with formal decision theory, a trained expert should act on their behalf to form a $LR$. As expounded throughout this paper, the interpretation of evidence in the form of a $LR$ is personal and subjective. We have not encountered any basis for the presumption that the surrogate $LR$ of an expert will reflect a truer implementation of decision theory than will the unquantified perception of the DM following effective presentation of the information upon which the expert's $LR$ is based.

Treating an expert's personal $LR$ as a quantity that is suitable for use in Bayes formula by others without modification gives it a status of *measurement* that it has not earned.[20] It is not traceable to any standard nor has it been calibrated across the pool of stakeholders. It does not come with an accompanying measure of uncertainty or reliability that could help the receiver of a $LR$ decide its fitness for purpose. Reproducibility properties of $LR$ characterizations, both exclusively among a pool of independent forensic experts and also inclusive of would-be DMs, are often overlooked. For a quantity defined as the ratio of two *subjective* or *personal* probabilities, the prospect that such investigations would yield substantial variability seems very real. Yet, the proposed modification of Lindley's framework uses the expert's $LR$ as though it were a perfect measurement of the DM's own weight of evidence.

In contrast, we believe that an expert's public characterization of weight of evidence, if given at all, should be modest, transparent and self-conscious in recognition that interpretation is personal.[21] One should openly convey the subjective, rather than normative, nature of choices made along the path taken from observations and experiences leading to the offered interpretation – for example, defining a relevant population, presuming available data to be representative (in the absence of rigorous probabilistic sampling), and selecting likelihood functions and prior distributions. Validation efforts should not be misrepresented to suggest that a particular model or narrow class of models is correct, rather than reasonable.[22] Whether a particular plausibility criterion and corresponding range of obtainable $LR$ values qualify the offered interpretation as acceptable is also a matter of personal opinion. In most cases, the range of $LR$ values attainable from models that would pass a given plausibility test is unknown, unless directly and rigorously pursued.

The dependence of probabilistic interpretations and any associated uncertainties upon modeling assumptions can be explored by conducting multiple analyses attempting to span the space of assumptions meeting a specified plausibility criterion. Presenting an uncertainty pyramid along with an explanation of the corresponding plausibility criterion and a description of the data may provide the audience the opportunity for greater understanding of the interactions between data, assumptions and interpretation.

If such uncertainty characterizations are considered untenable, one may be forced to conclude that the adapted Lindley plan, though appealing in theory, is impractical to implement. It does not mean that, just because one is unable to calculate the required value, one should accept the value that can be calculated. Alternatively, an $LR$ value may still be transferable as a discriminant score attempting to sort between realizations under $H_p$ and $H_d$. From this perspective a $LR$ value would not have direct, probabilistic interpretation, but its meaning would become unmasked through empirical evaluations of its performance.[23] More broadly, objective descriptions of procedures followed and outcomes obtained throughout investigation of the case and broader experience may present a promising path to ensuring transferability of information from a forensic expert to DMs.

We hope the arguments presented throughout this paper will encourage the forensic science community to be mindful of the many components of uncertainty that become evident when viewing a $LR$ as a measurement. Correspondingly, we hope best-practice guidances address how to avoid overstating the authority or rigor underlying any particular interpretation of evidence.

---

[20] See JCGM (2008) and Possolo (2015) for a discussion of what constitutes a measurement.

[21] "I emphasize that the answers you give to the questions I ask you about your uncertainty are yours alone, and need not be the same as what someone else would say, even someone with the same information as you have, and facing the same decisions. " – Kadane (2011)

[22] As a result of this common misunderstanding, we prefer phrases that use "plausible" or "plausibility" in place of "validated" or "validation", respectively.

[23] Receiver operating characteristic (ROC) curves (Peterson and Birdsall, 1953) are invaluable in such evaluations.

Additionally, we hope the community pursues tools for descriptive presentations intended to assist the DMs in directly establishing their own respective interpretations of evidence.

**Acknowledgments:**

The authors would like to acknowledge William Guthrie, Dr. Martin Herman, Dr. Adam Pintar, Professor David Kaye, and Professor Jacqueline Speir for their comments and suggestions which were very helpful in making a number of improvements to the manuscript.

# Appendix

In this appendix we display the results for additional choices of $F$ and $G_0$. Six different choices are considered for $G_0$. They are explained below:

| | |
|---|---|
| Figure 6 | $t$ distribution with 1 df |
| Figure 7 | $t$ distribution with 0.14 df |
| Figure 8 | $\chi^2$ distribution with 3 df |
| Figure 9 | $\chi^2$ distribution with 2 df |
| Figures 10-12 | Symmetric unimodal nonparametric distributions |
| Figures 13-15 | Unimodal nonparametric distributions |
| Figures 16-18 | Symmetric Nonparametric distributions |
| Figures 19-21 | Nonparametric distributions |

The top plot in each figure shows the 95% Kolmogorov-Smirnov confidence band for the $CDF$ of refractive indices from 49 fragments from a single window (Bennett Data). The empirical $CDF$ is shown in gray. The faded red and green smooth curves, respectively, correspond to members of the chosen scale family with the smallest and largest scaling factors such that, the discrete distributions obtained by accounting for interval-censoring in the reported data (shown using solid red and solid green line segments, respectively), are entirely contained within the confidence band.

The bottom plot in each figure displays the $LR$ values corresponding to various choices for $F$, reflected by position along the x-axis, and the scale factor used with the shape chosen for $G_0$. The left-most results correspond to the estimate of $F$ labeled as Jump, which is displayed in Figure 2. The remaining positions reflect the bandwidth of the Gaussian kernel leading to the estimate of $F$ used in computing the $LR$. Within each choice of $F$, the $LR$ values are staggered in order of the scale parameter used with $G_0$ to emphasize the potential non-monotonic relationship between $LR$ and the scale parameter. The points are color-coded to indicate the associated scale parameter values in accordance with the legend titled $\sigma_{\text{within}}$.

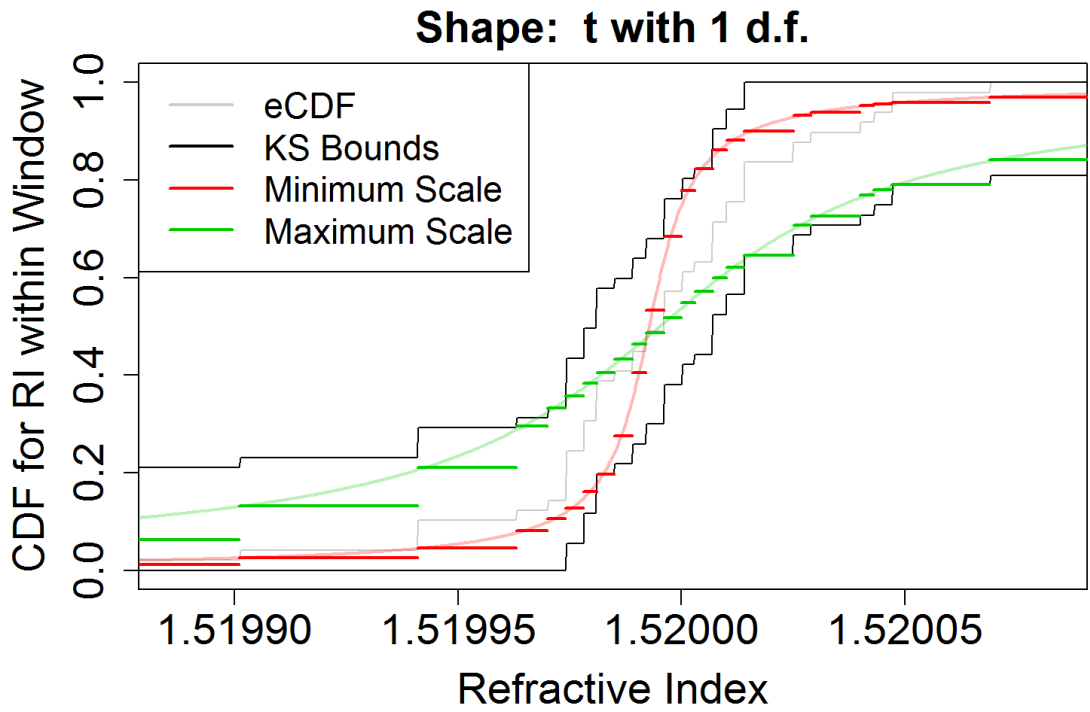
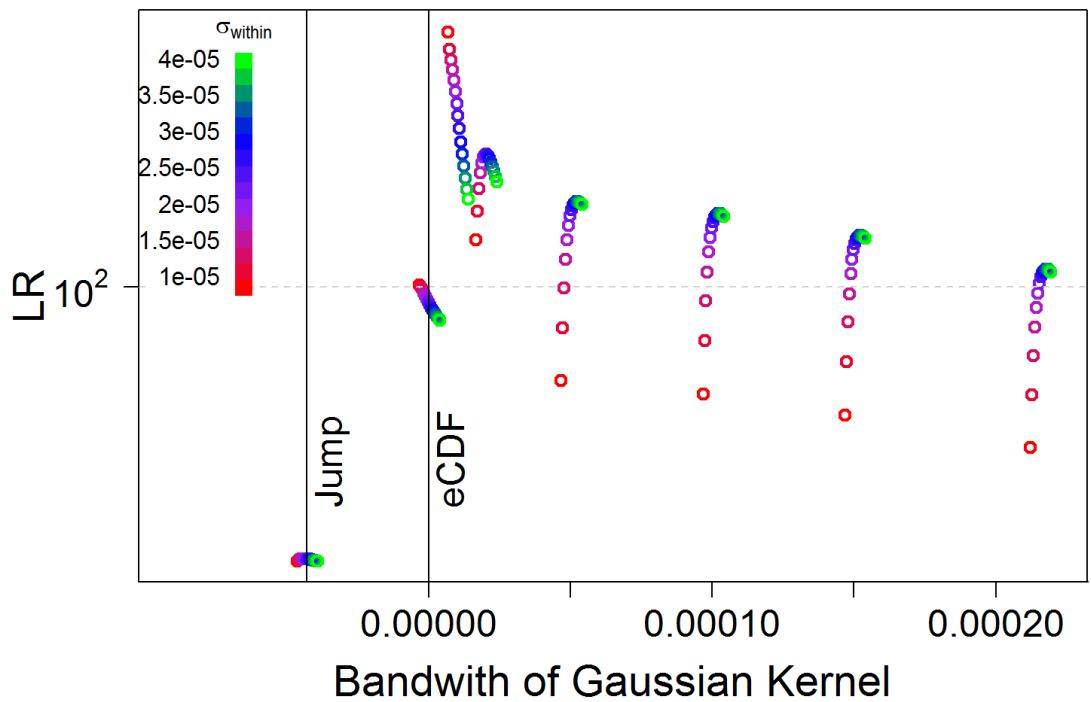

**Figure 6:** $LR$ values when $G_0$ is a $t$ distribution with 1 degree of freedom.

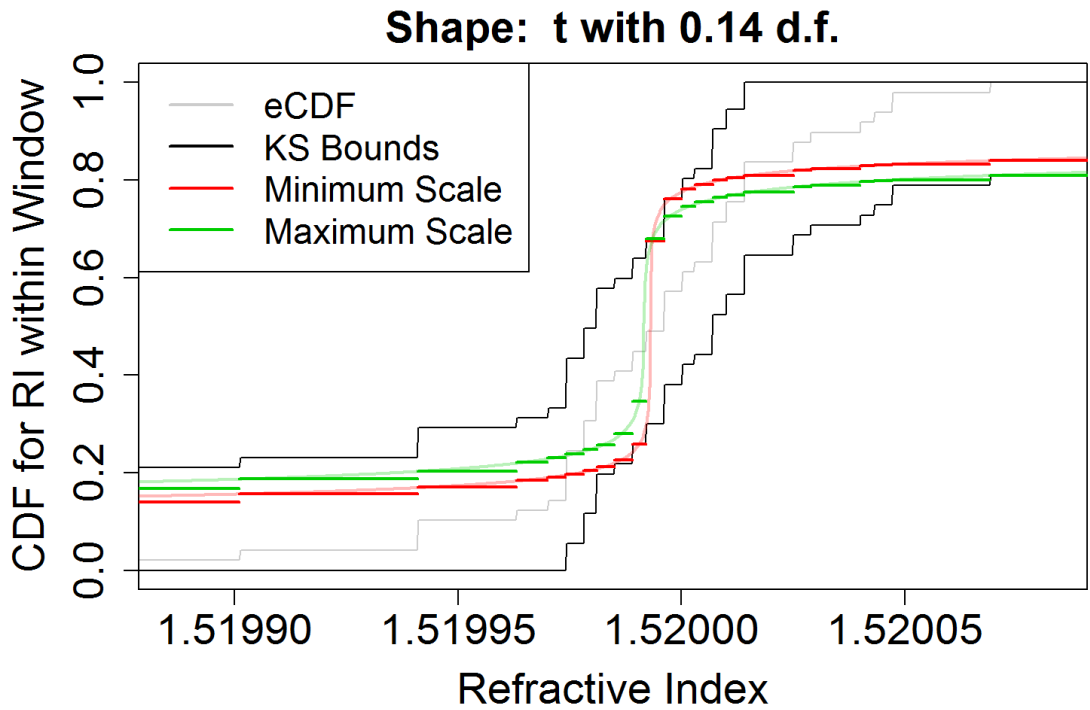
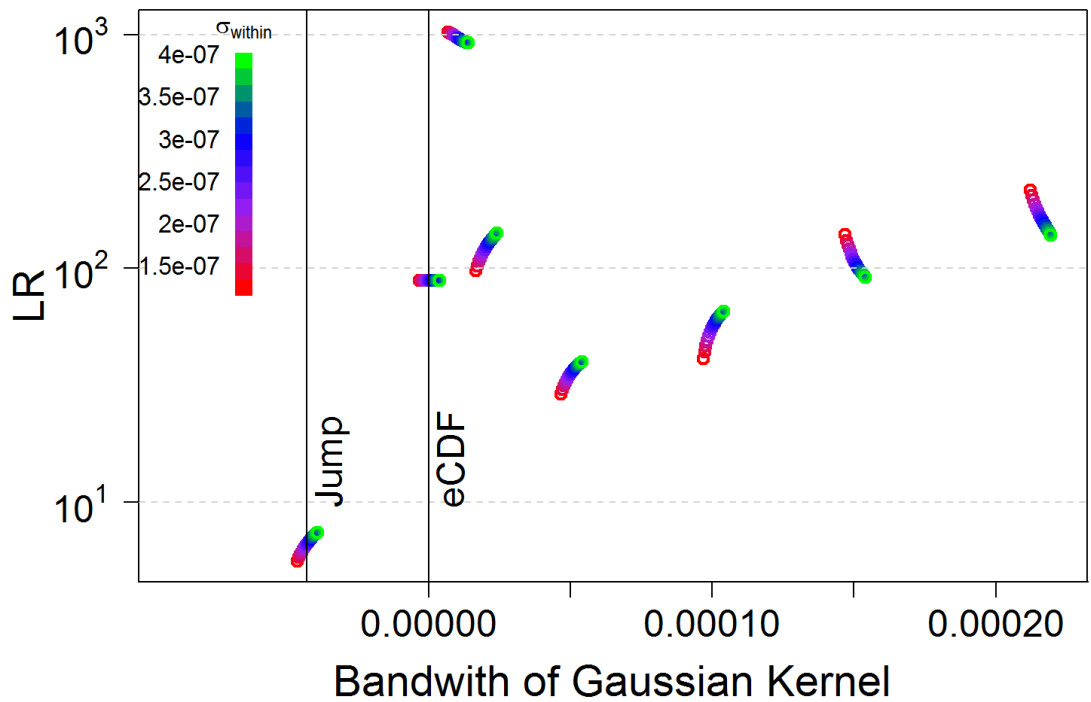

**Figure 7:** $LR$ values when $G_0$ is a $t$ distribution with 0.14 degree of freedom.

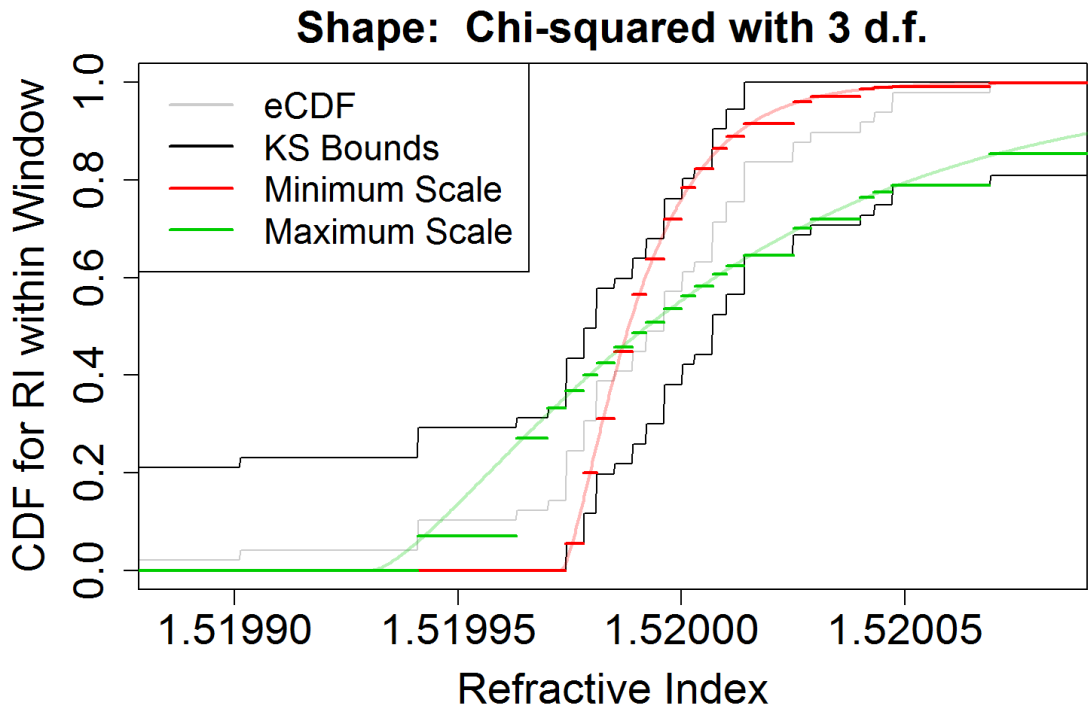

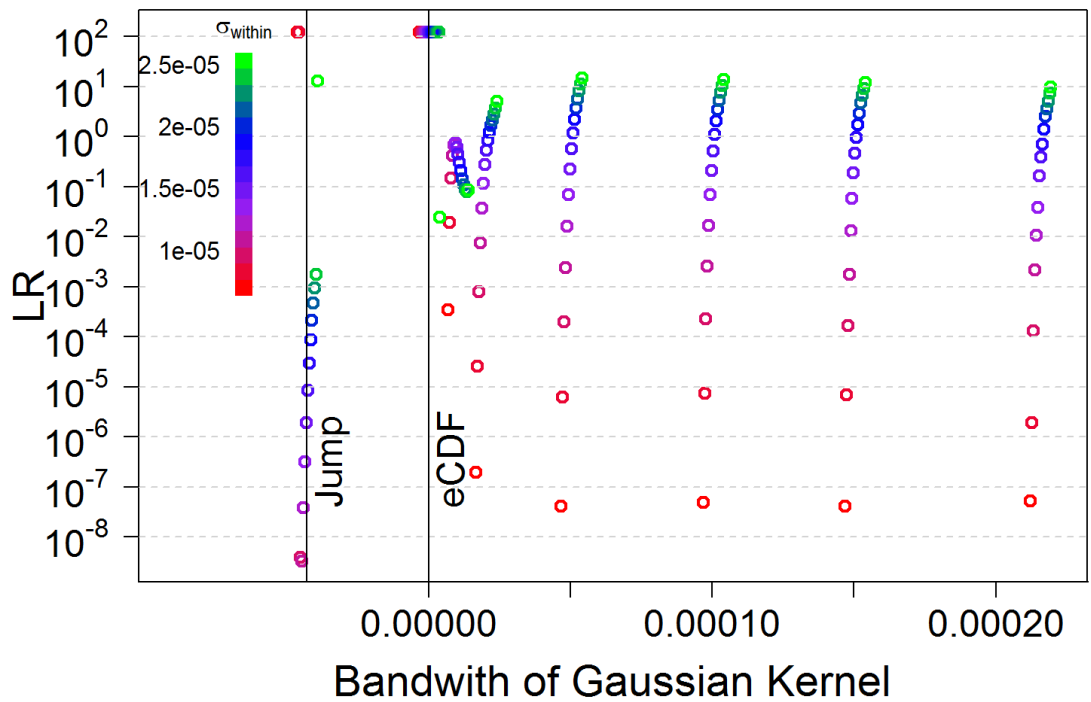

**Figure 8:** $LR$ values when $G_0$ is a $\chi^2$ distribution with 3 degrees of freedom.

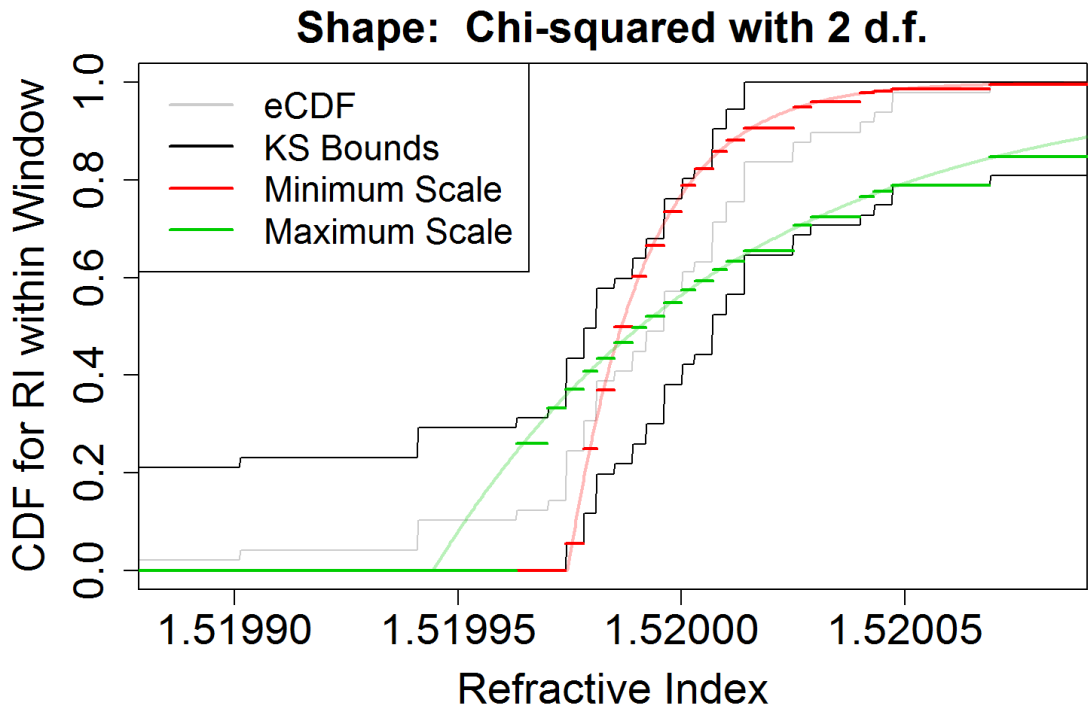

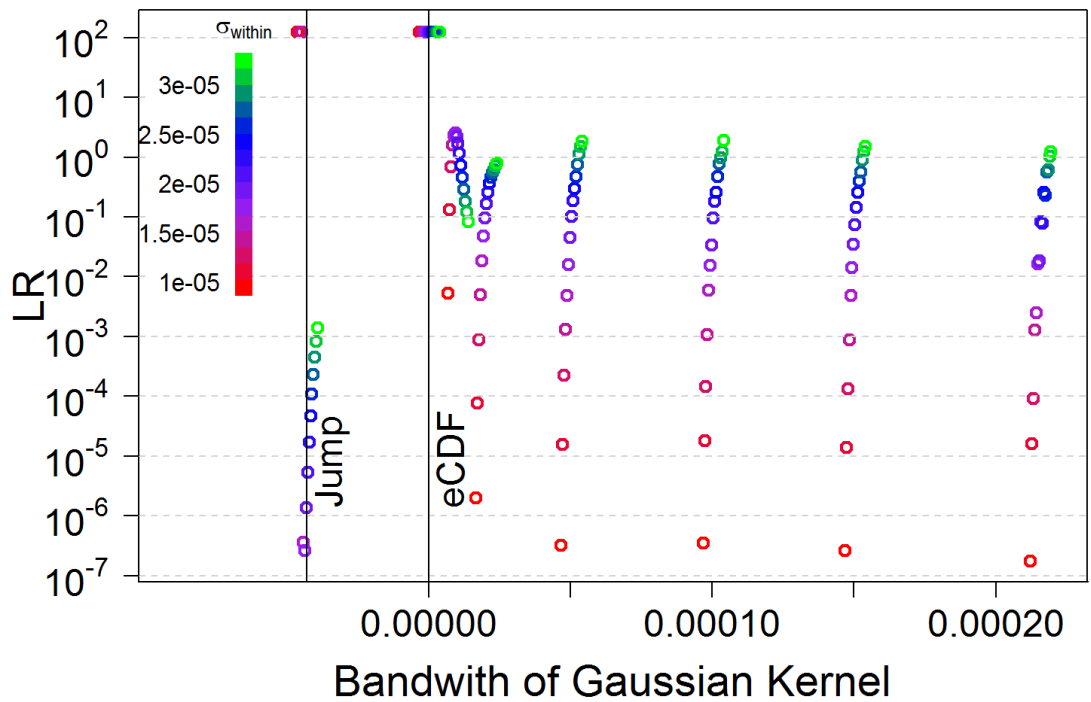

**Figure 9:** $LR$ values when $G_0$ is a $\chi^2$ distribution with 2 degrees of freedom.

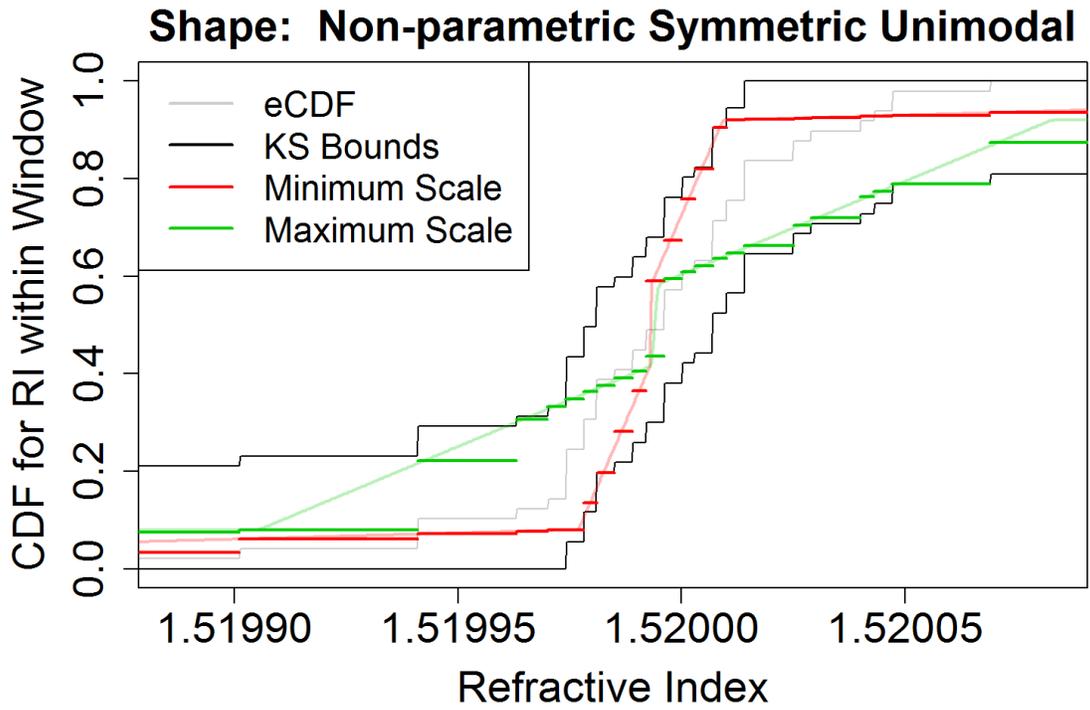
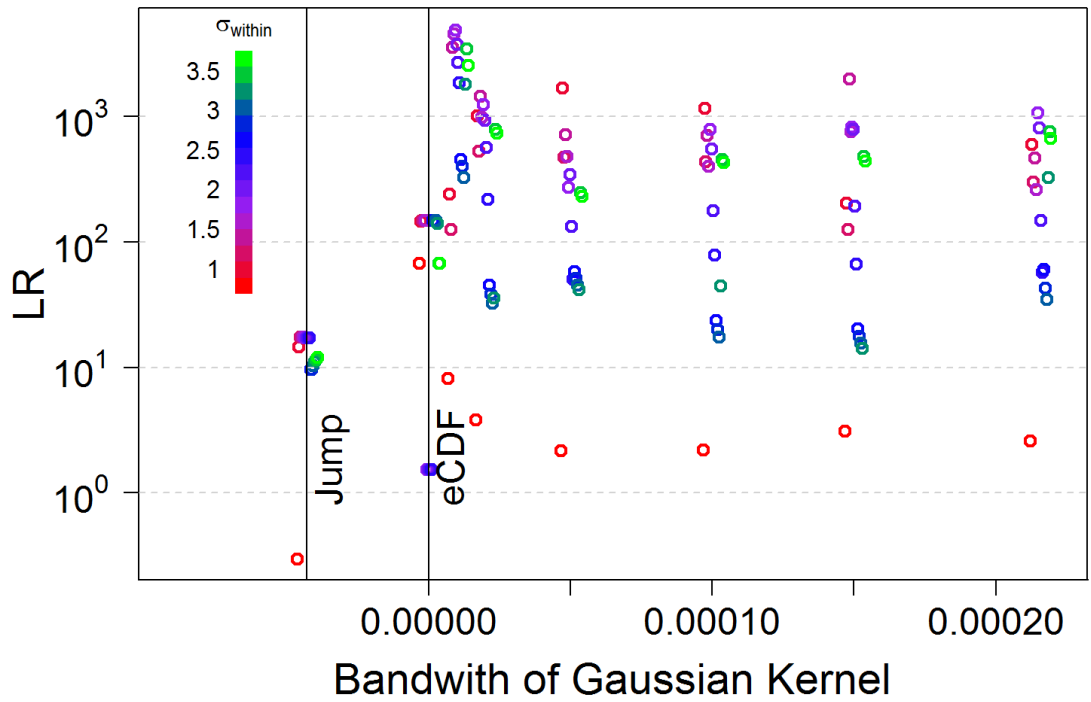

**Figure 10:** $LR$ values when $G_0$ has the shape of the nonparametric symmetric unimodal distribution shown.

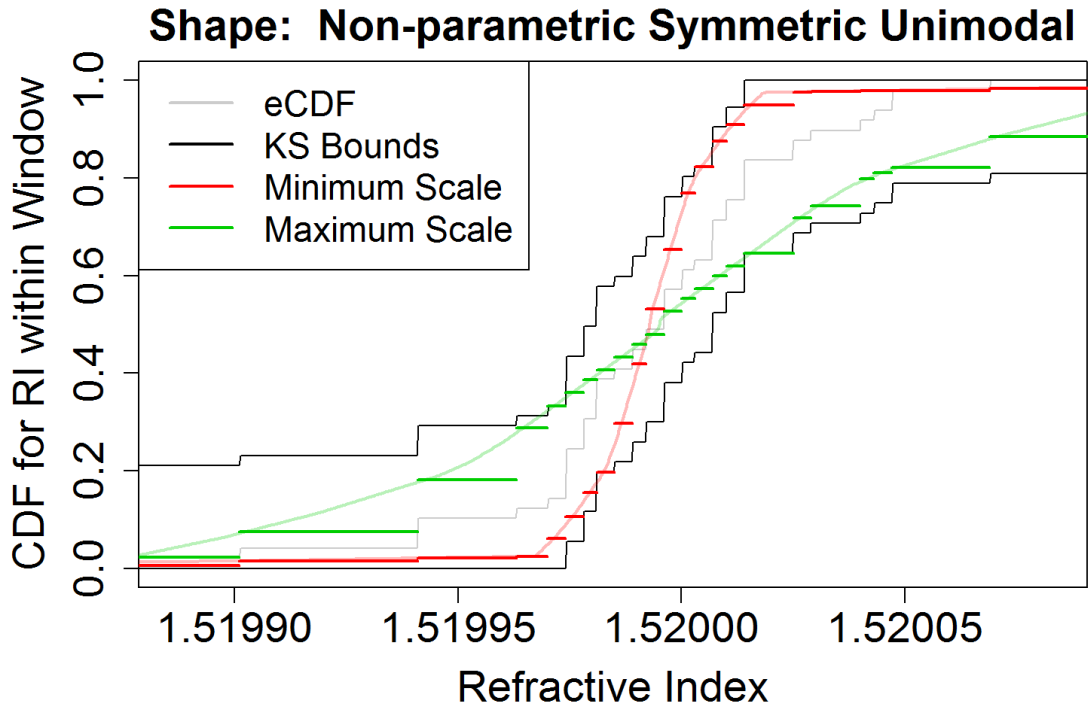

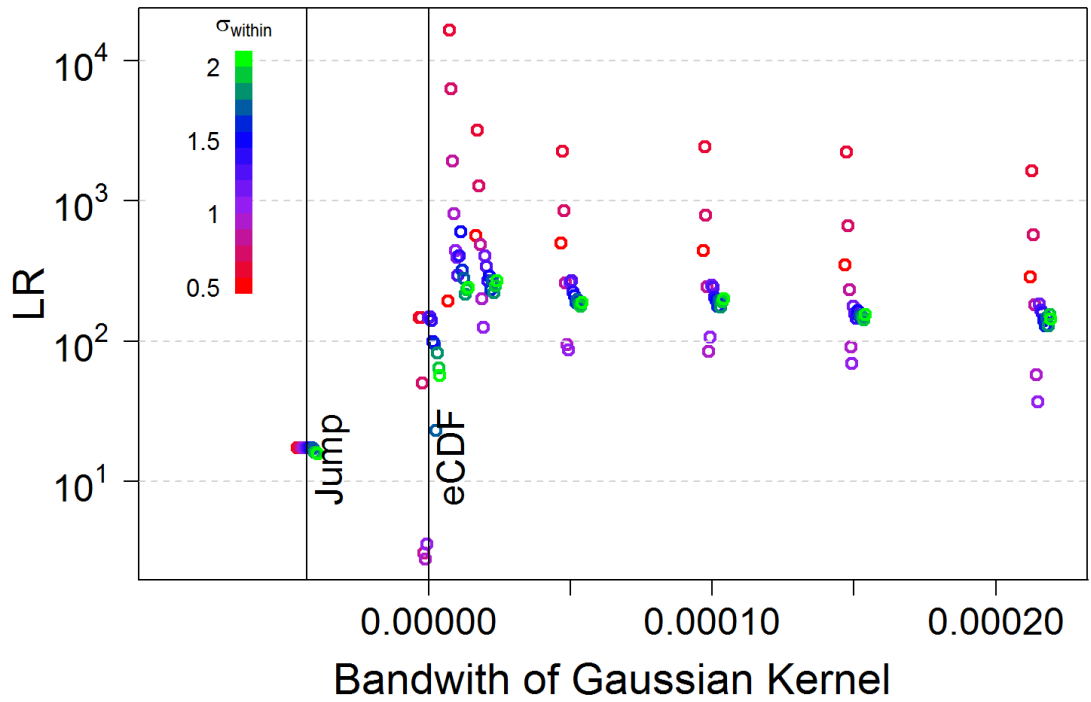

**Figure 11:** $LR$ values when $G_0$ has the shape of the nonparametric symmetric unimodal distribution shown.

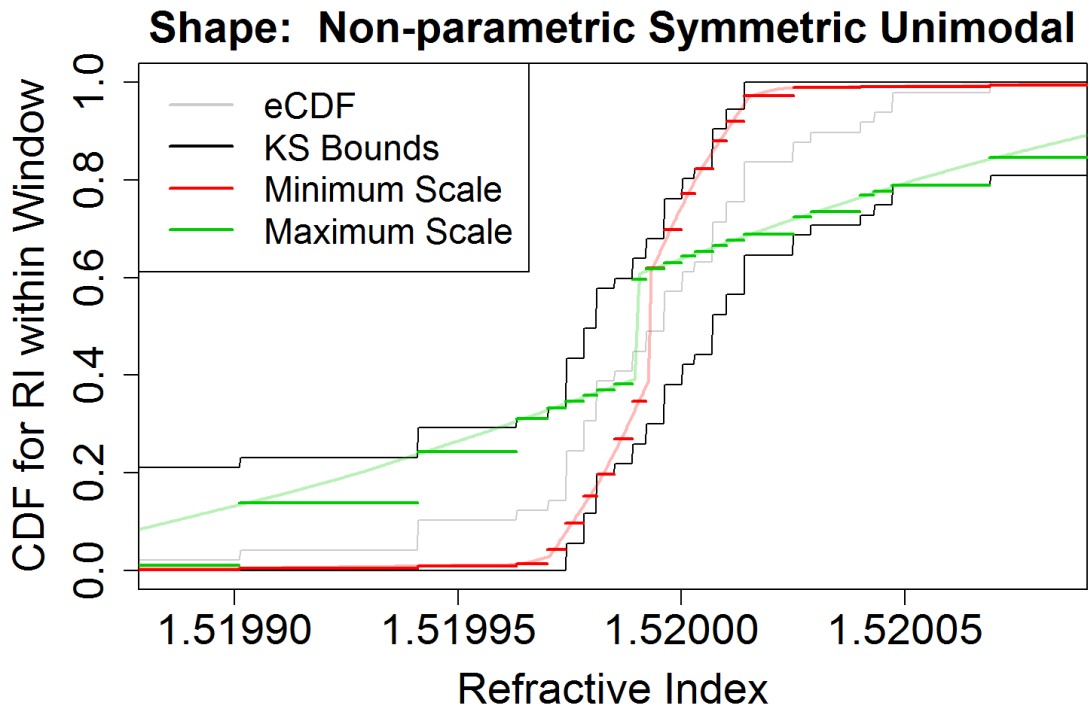

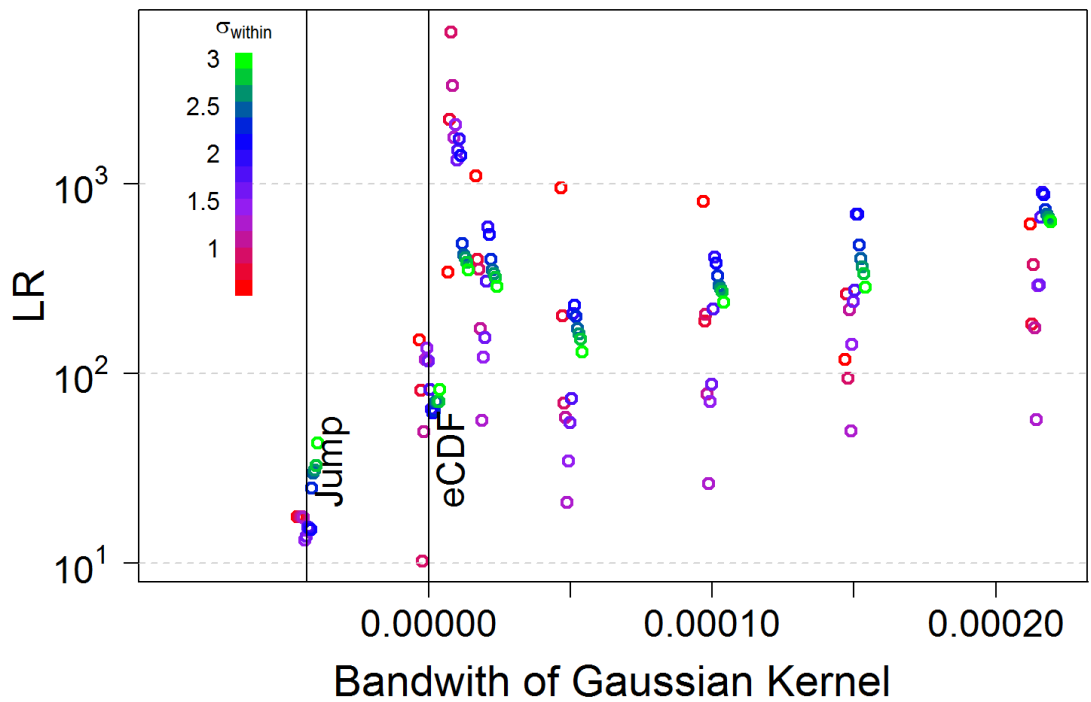

**Figure 12:** $LR$ values when $G_0$ has the shape of the nonparametric symmetric unimodal distribution shown.

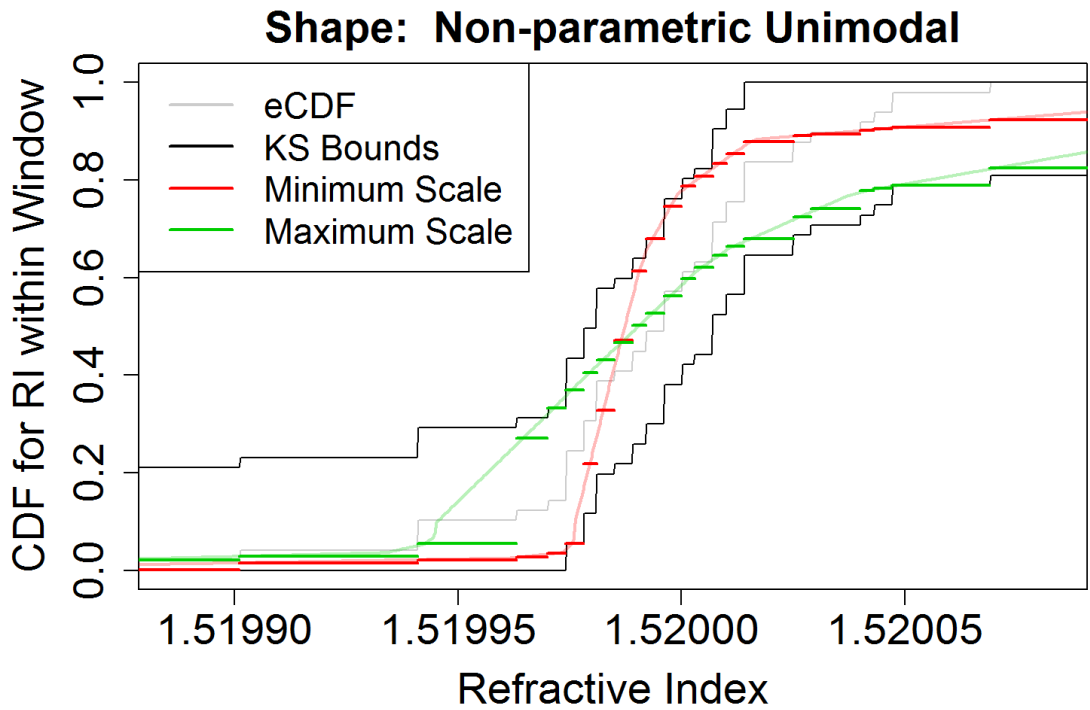
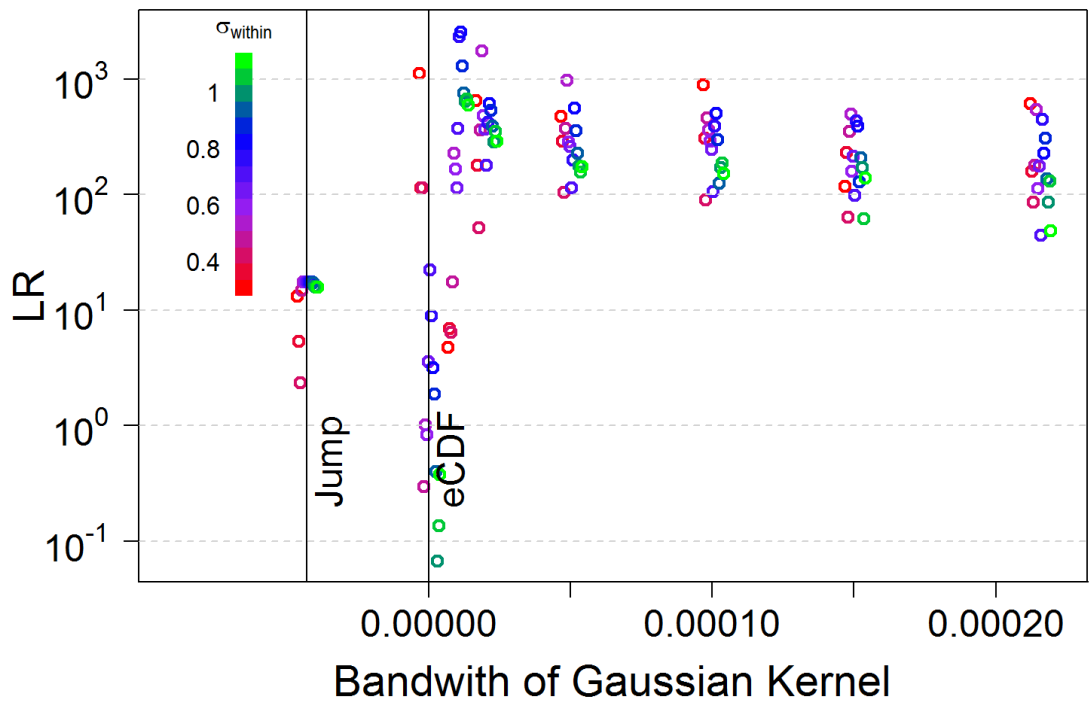

**Figure 13:** $LR$ values when $G_0$ has the shape of the nonparametric unimodal distribution shown.

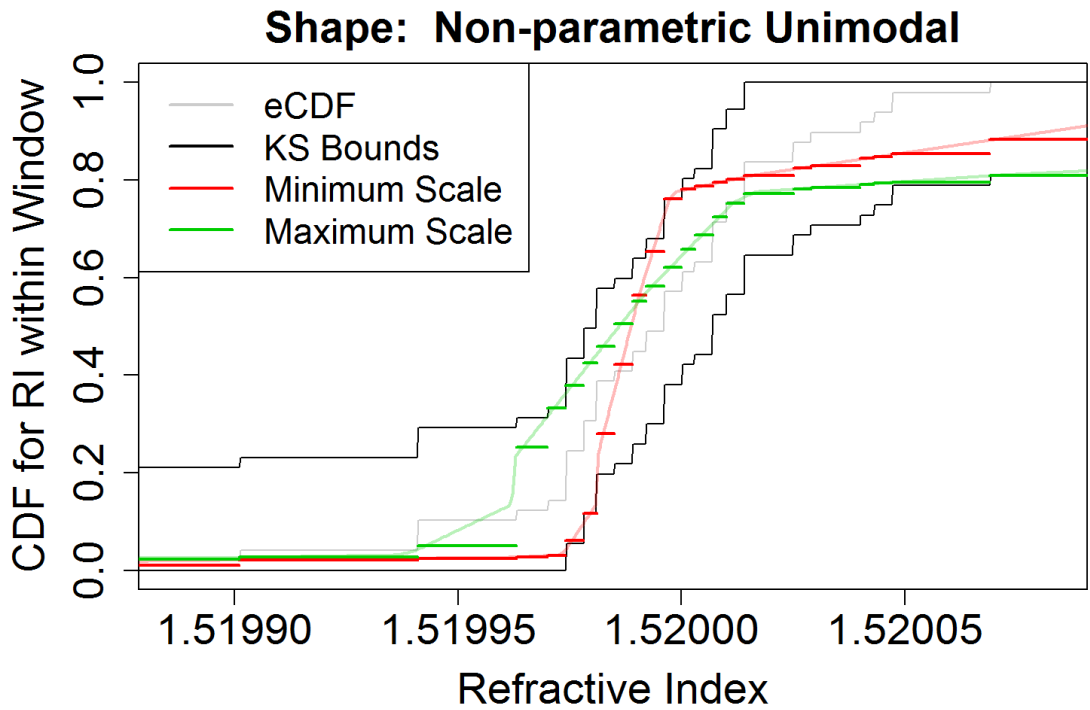
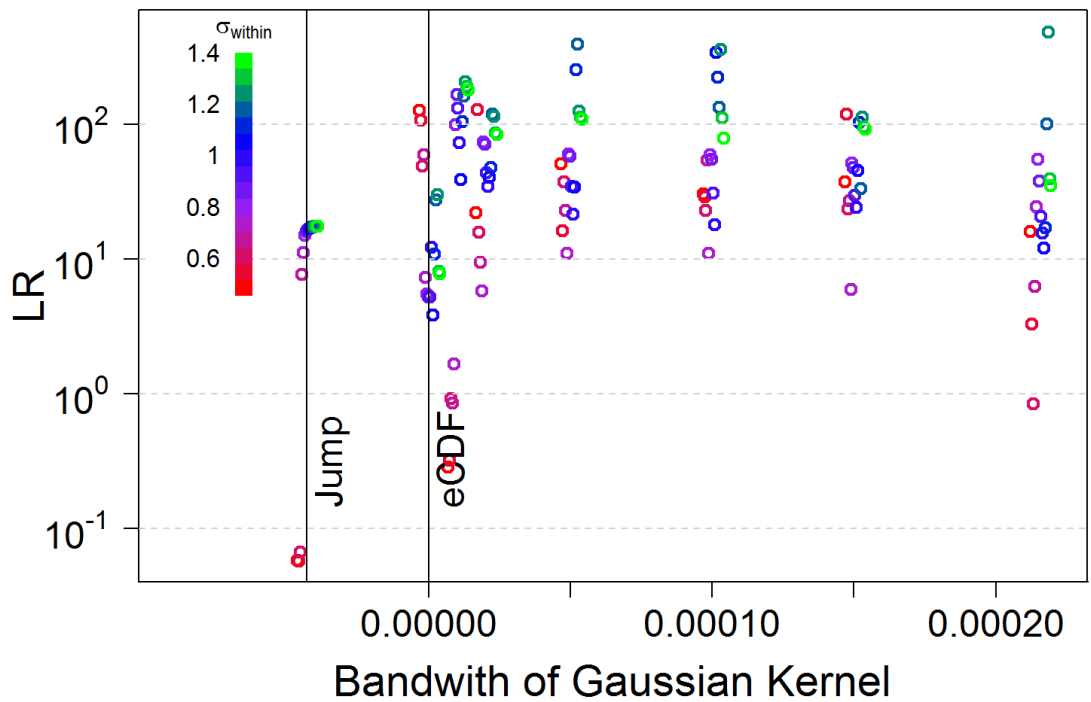

**Figure 14:** $LR$ values when $G_0$ has the shape of the nonparametric unimodal distribution shown.

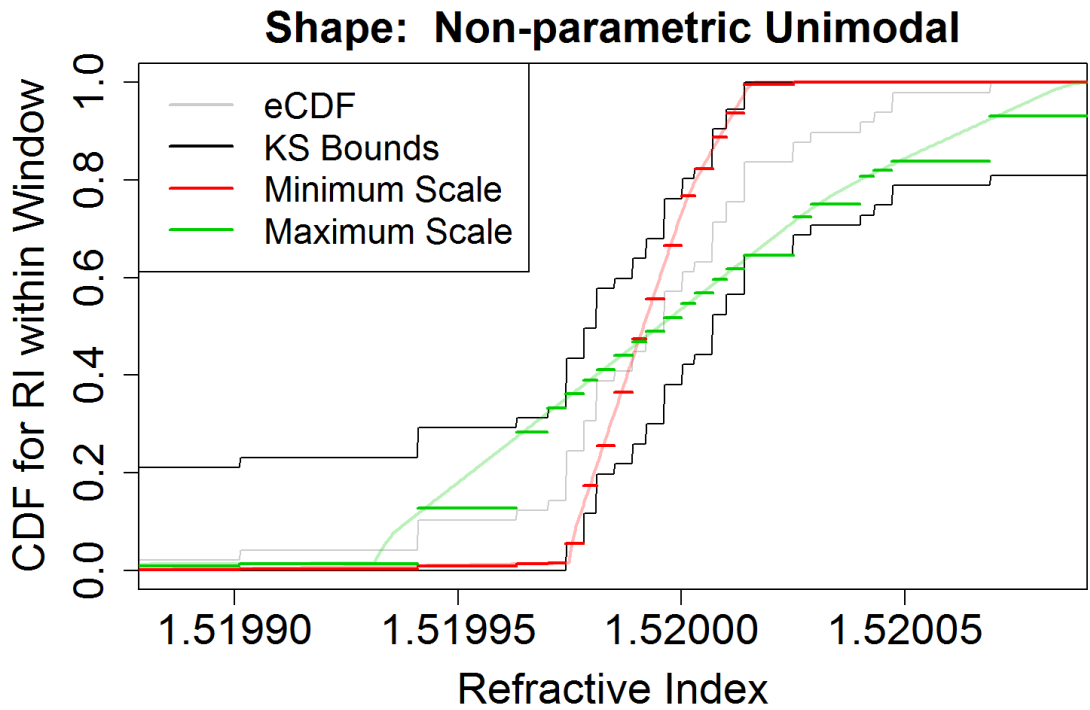

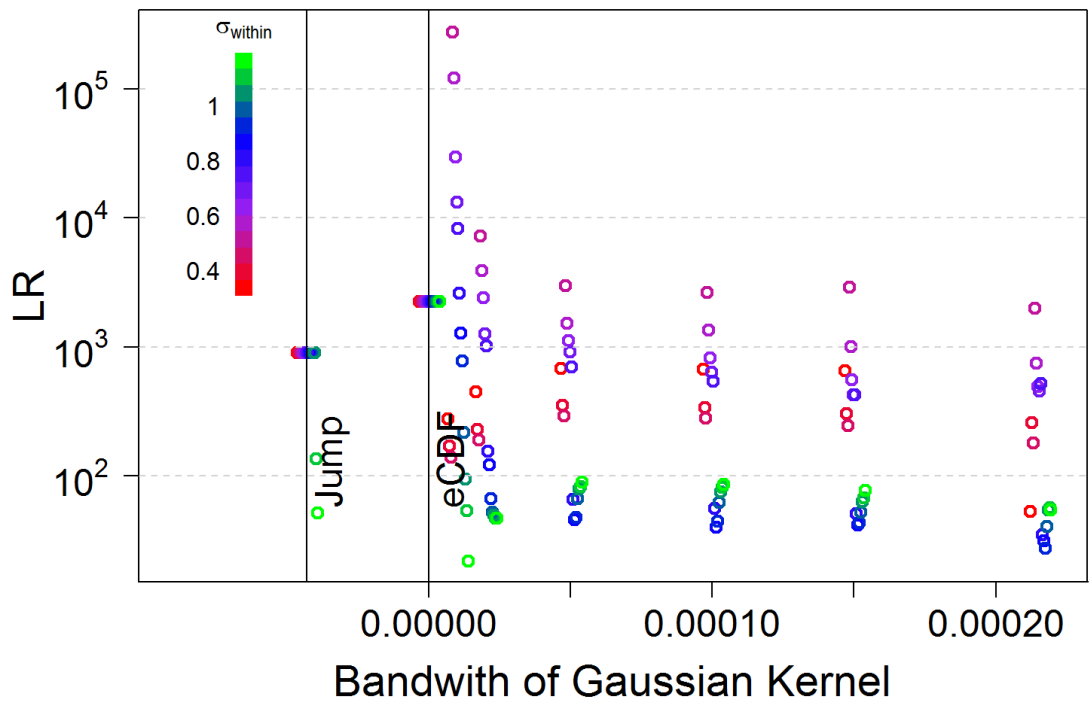

**Figure 15:** $LR$ values when $G_0$ has the shape of the nonparametric unimodal distribution shown.

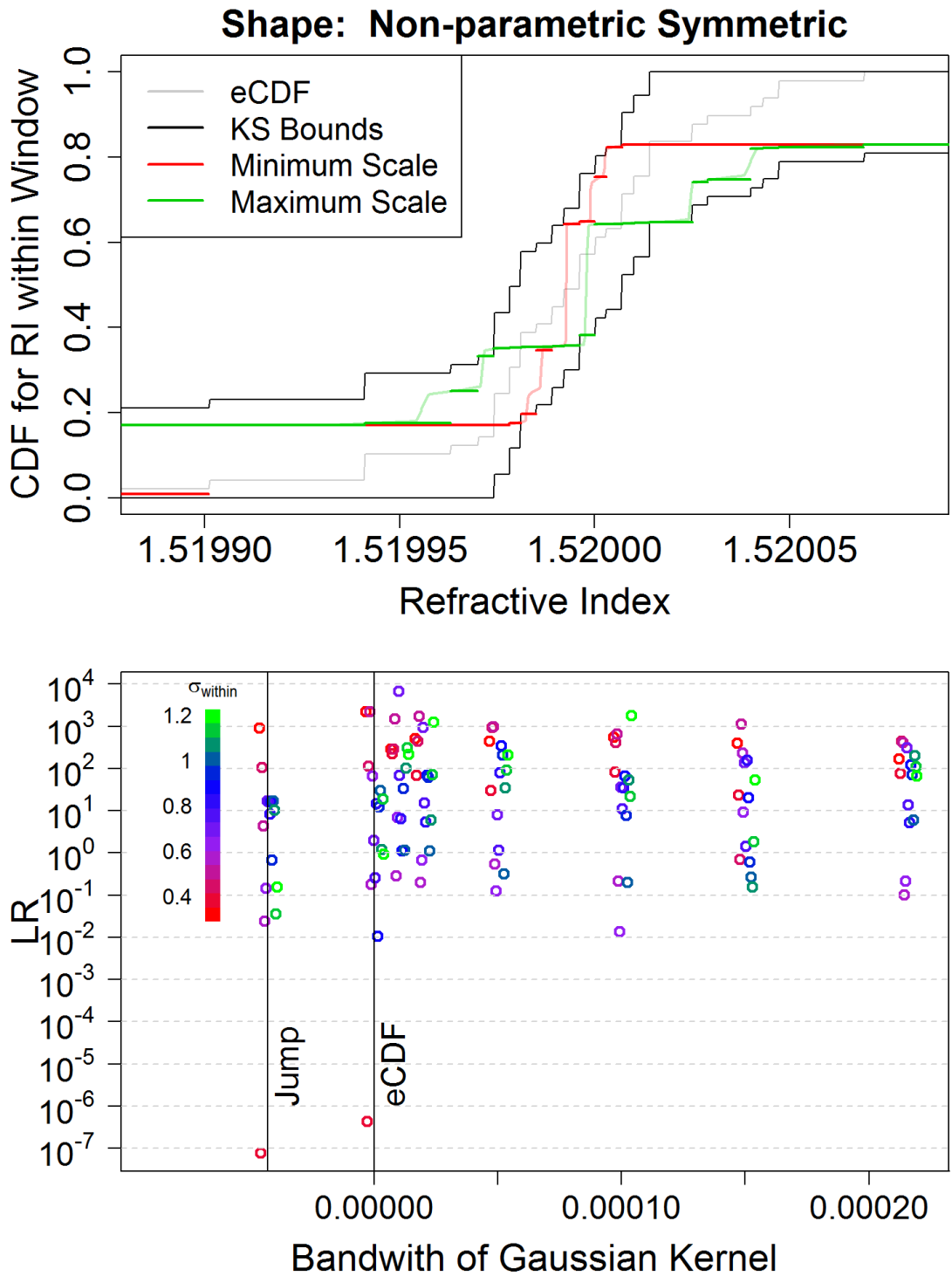

**Figure 16:** $LR$ values when $G_0$ has the shape of the nonparametric symmetric distribution shown.

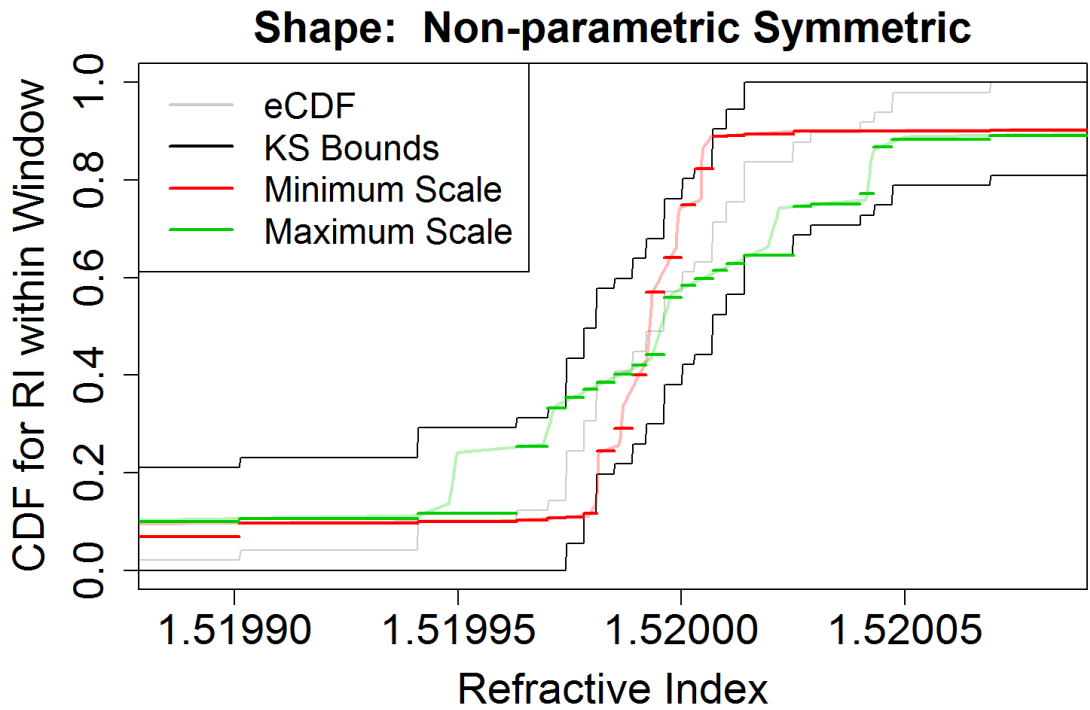

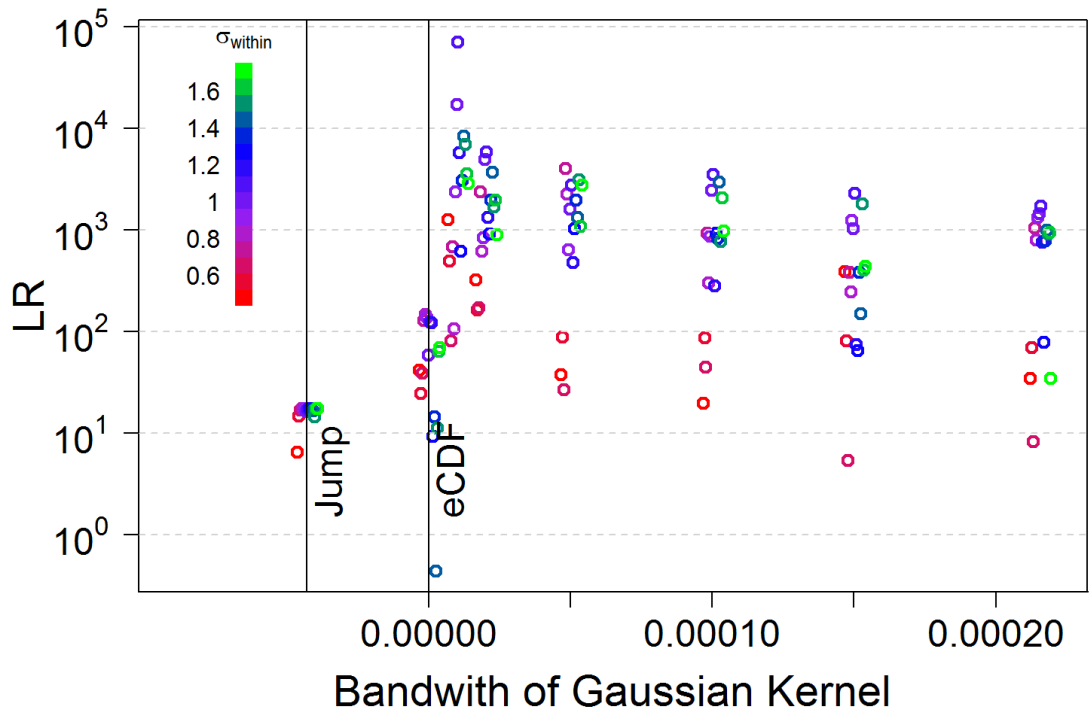

**Figure 17:** $LR$ values when $G_0$ has the shape of the nonparametric symmetric distribution shown.

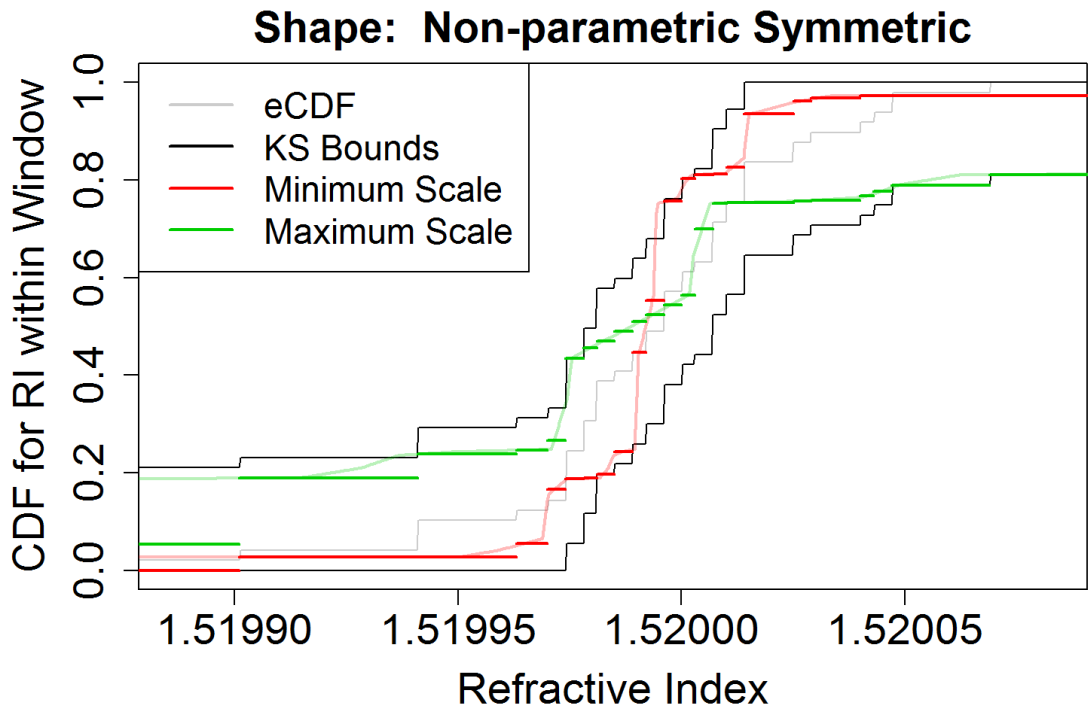

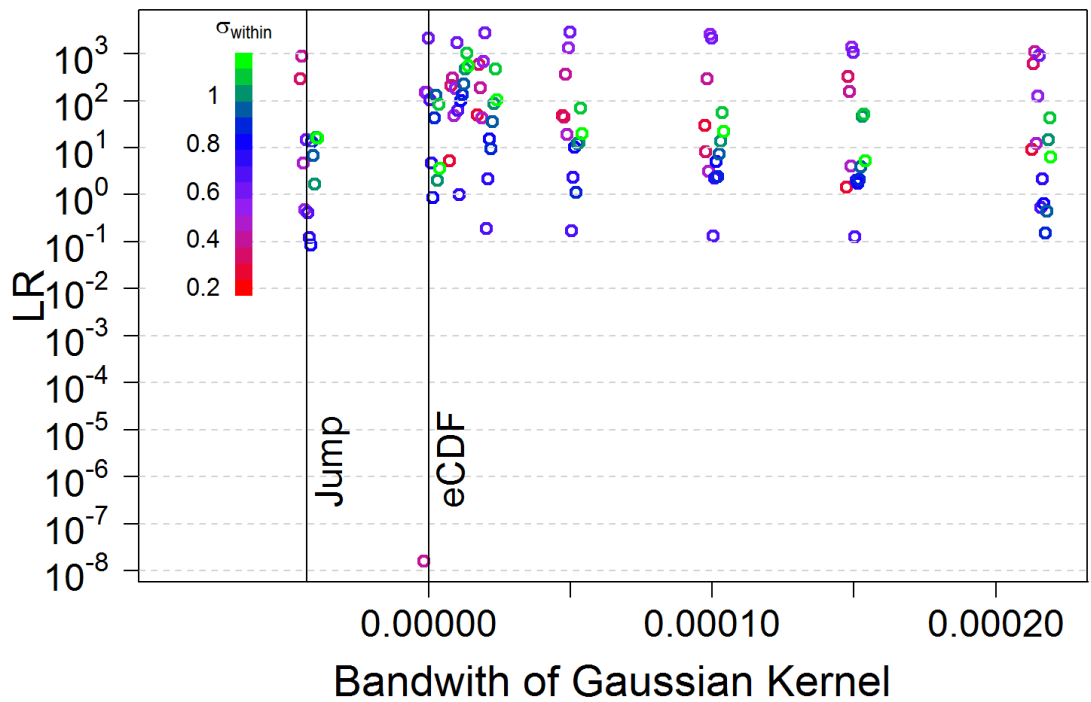

**Figure 18:** $LR$ values when $G_0$ has the shape of the nonparametric symmetric distribution shown.

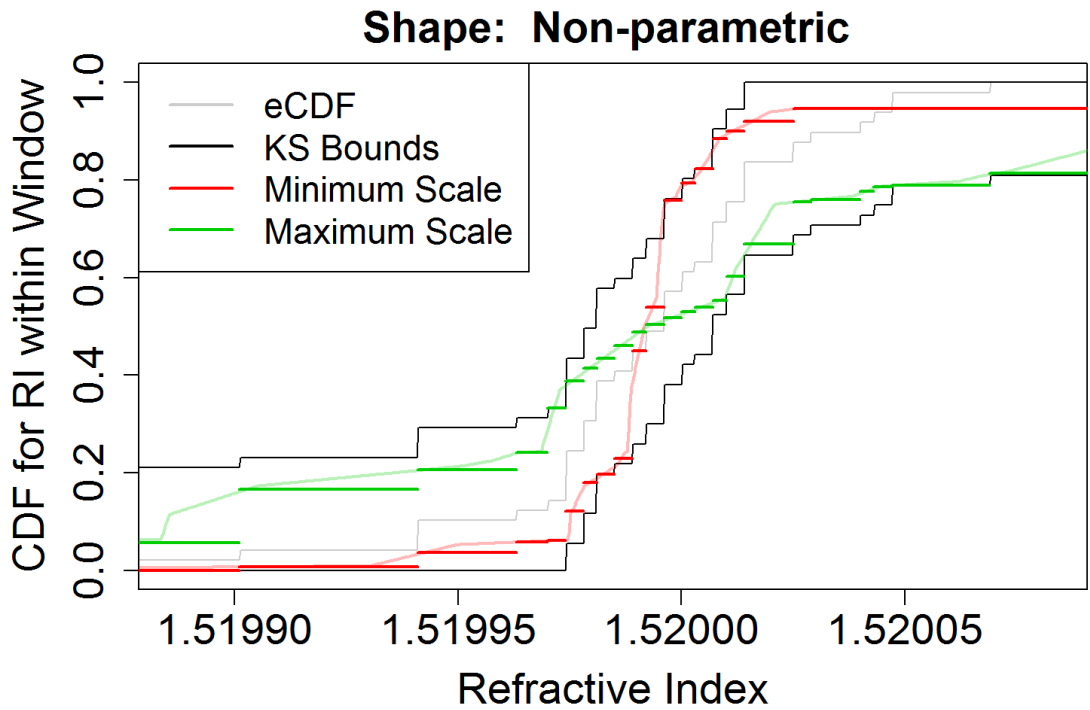

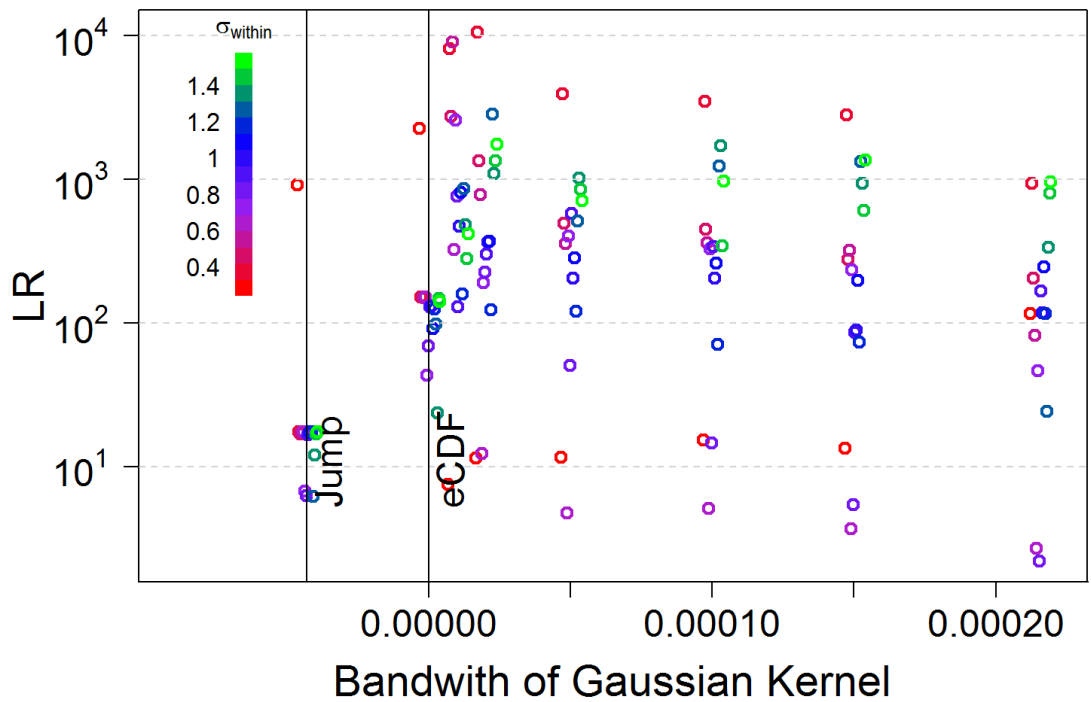

**Figure 19:** $LR$ values when $G_0$ has the shape of the nonparametric distribution shown.

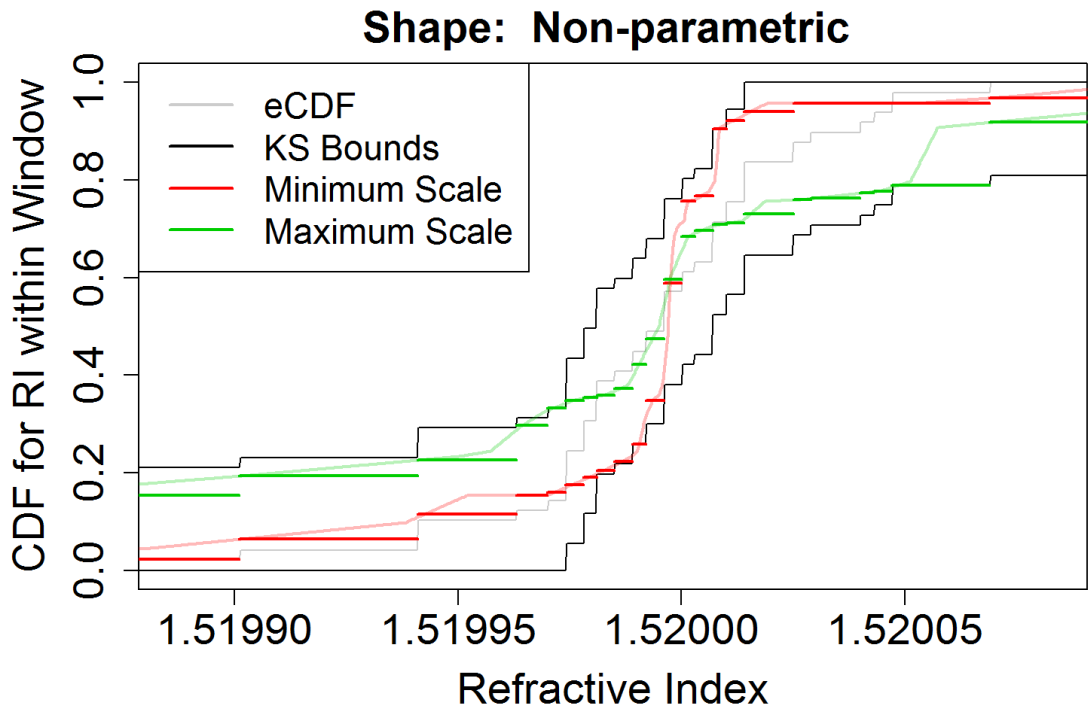

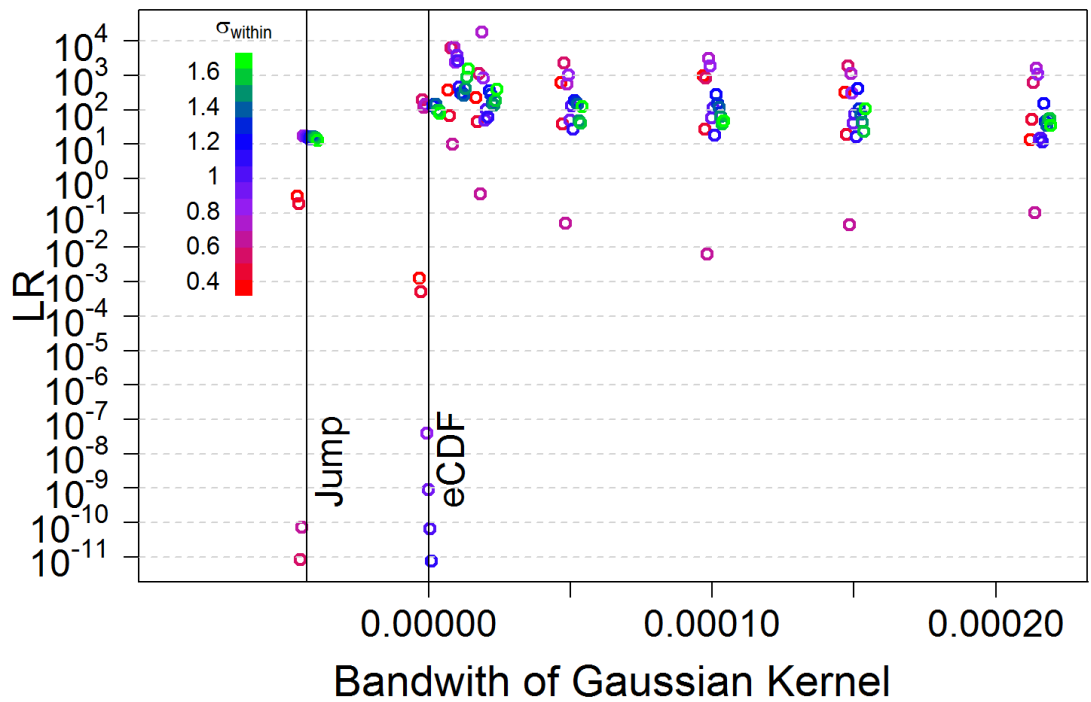

**Figure 20:** $LR$ values when $G_0$ has the shape of the nonparametric distribution shown.

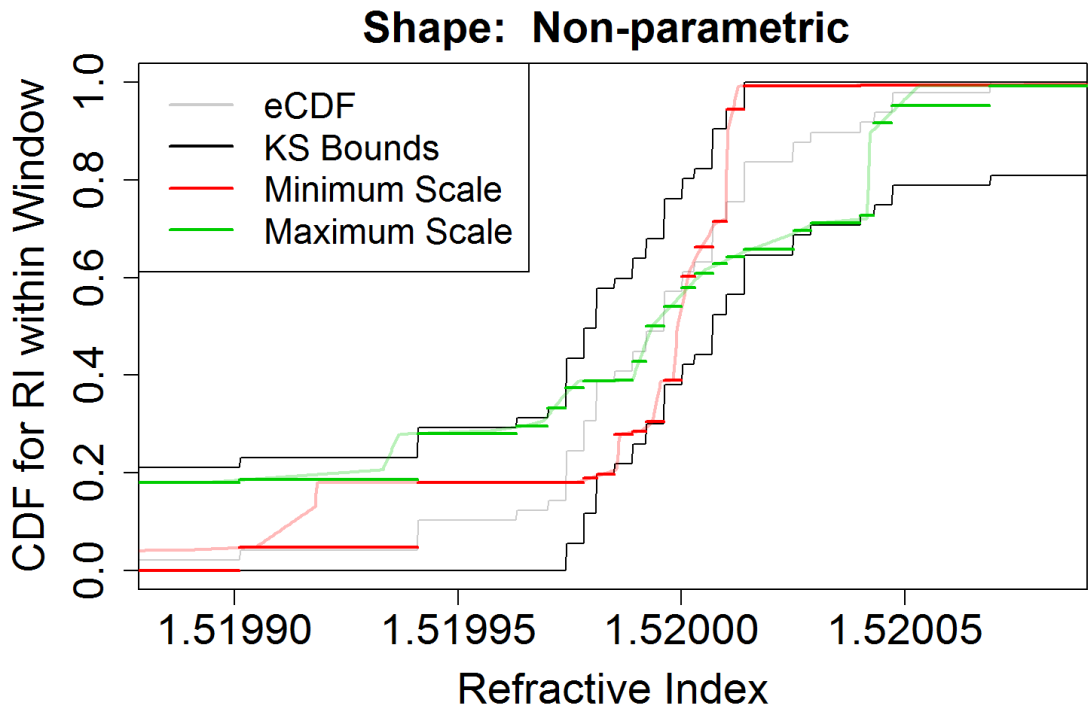

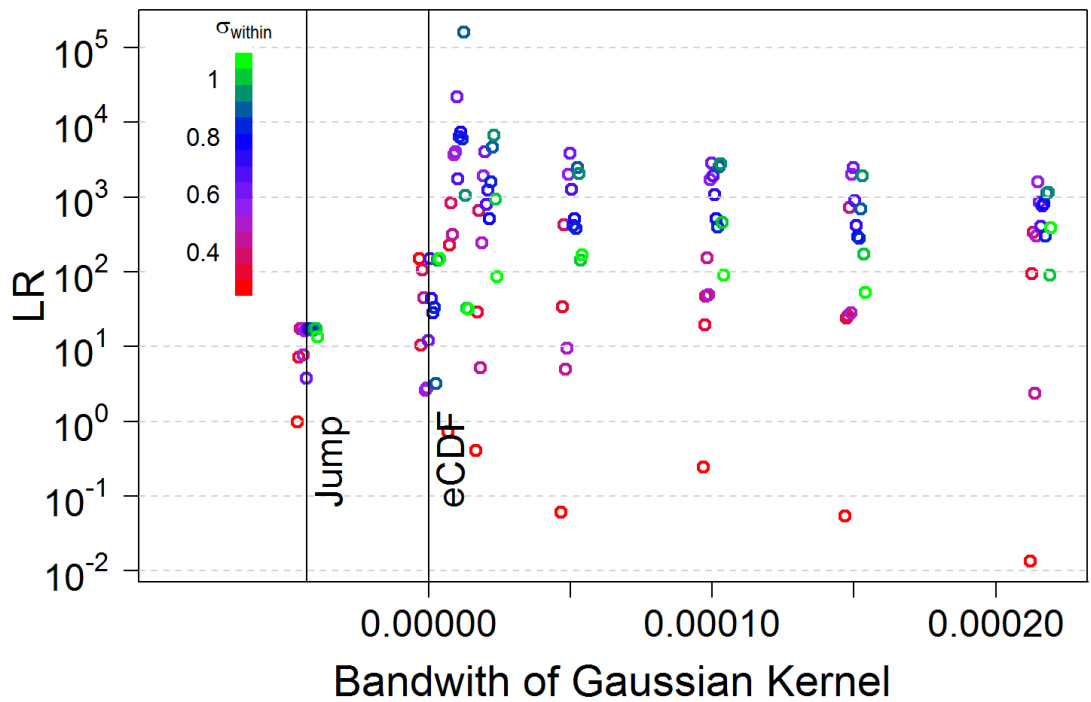

**Figure 21:** $LR$ values when $G_0$ has the shape of the nonparametric distribution shown.